\documentclass[
 reprint,
 superscriptaddress,
 noshowpacs, 
 noshowkeys, 
 amsmath, amssymb,
 aps,
 prl
]{revtex4-1}

\usepackage{graphicx}
\usepackage{dcolumn}
\usepackage{bm}
\usepackage[colorlinks=true, allcolors=blue]{hyperref}
\usepackage{siunitx}
    \DeclareSIUnit \samples{S}
    \DeclareSIUnit \counts{Cts}
    \DeclareSIUnit \gauss{G}
\begin{document}


\title{Near-Infrared-Assisted Charge Control and Spin Readout\\ of the Nitrogen-Vacancy Center in Diamond}


\author{David A.\ Hopper}
	\affiliation{Quantum Engineering Laboratory, Department of Electrical and Systems Engineering, University of 			Pennsylvania, Philadelphia, PA 19104 USA}
    \affiliation{Department of Physics, University of Pennsylvania, Philadelphia, PA 19104 USA}
\author{Richard R. Grote}
	\affiliation{Quantum Engineering Laboratory, Department of Electrical and Systems Engineering, University of 			Pennsylvania, Philadelphia, PA 19104 USA}
\author{Annemarie L.\ Exarhos}
	\affiliation{Quantum Engineering Laboratory, Department of Electrical and Systems Engineering, University of 			Pennsylvania, Philadelphia, PA 19104 USA}
\author{Lee C.\ Bassett}
	\email[Corresponding author.\\ Email address: ]{lbassett@seas.upenn.edu}
	\affiliation{Quantum Engineering Laboratory, Department of Electrical and Systems Engineering, University of 			Pennsylvania, Philadelphia, PA 19104 USA}
\date{\today}

\begin{abstract}
We utilize nonlinear absorption to design all-optical protocols that improve both charge state initialization and spin readout for the nitrogen-vacancy (NV) center in diamond.  Non-monotonic variations in the equilibrium charge state as a function of visible and near-infrared optical power are attributed to competing multiphoton absorption processes. In certain regimes, multicolor illumination enhances the steady-state population of the NV's negative charge state above 90\%.  At higher NIR intensities, selective ionization of the singlet manifold facilitates a protocol for spin-to-charge conversion that dramatically enhances the spin readout fidelity.  We demonstrate a 6-fold increase in the signal-to-noise ratio for single-shot spin measurements and predict an orders-of-magnitude experimental speedup over traditional methods for emerging applications in magnetometry and quantum information science using NV spins.
\end{abstract}

\maketitle


The diamond nitrogen-vacancy (NV) center is a versatile solid-state qubit \cite{Awschalom2013} and nanoscale sensor \cite{Lovchinsky2016,Karaveli2016}, exhibiting long spin coherence times at room temperature and all-optical mechanisms for qubit initialization and readout. Unfortunately, these convenient mechanisms are imperfect.  High-fidelity qubit initialization and readout are crucial requirements for large-scale quantum information processing \cite{DiVincenzo2000}, but the NV's intrinsic optical dynamics limit the charge and spin initialization purity well below unity \cite{Waldherr2011,Waldherr2011a,Robledo2011a,Aslam2013} and only provide a low-fidelity spin-state readout \cite{Taylor2008,Jiang2009,Lovchinsky2016}.  These drawbacks impose substantial averaging requirements that diminish the full potential of the NV center as a qubit and quantum sensor.

Initialization and readout of NV spins are traditionally achieved through optical excitation and photoluminescence (PL) detection, respectively.  Optical excitation, however, causes unavoidable cycling from the desired negative charge state (NV$^-$) into the neutral state (NV$^0$) with a different energy structure. Charge-state transitions result from optical excitation of an electron from NV$^-$ to the conduction band (ionization) and a hole from NV$^0$ to the valence band (recombination), which compete to produce a maximum steady-state NV$^-$ population of $\sim75\%$ using an optimized single excitation wavelength of \SI{532}{\nano\meter} \cite{Aslam2013}.

Traditional PL-based spin readout results from a spin-dependent inter-system crossing (ISC) between the triplet and singlet manifolds of NV$^-$ \cite{Oort1988, Robledo2011a,Doherty2013}. In typical experiments, PL contrast only exists for the first $\sim$\SI{200}{\nano\second} of excitation, during which time $\sim$0.01 PL photons are collected, on average. Therefore $\sim10^4$ repeats are required to obtain adequate signal-to-noise ratio (SNR). This averaging requirement precludes important applications including projective \cite{Pfaff2013} and partial \cite{Blok2014} measurements of proximal nuclear spins, as well as verification of entanglement between remote NVs \cite{Bernien2013}, all of which rely on a single-shot electron readout protocol that, at present, is only available at cryogenic temperatures \cite{Robledo2011}.

Here, we demonstrate dramatic improvements in both charge initialization and spin readout fidelity by using multicolor illumination to manipulate the NV's spin and charge-state dynamics. Our experiments uncover complex multiphoton absorption effects driven by visible and near-infrared excitation and resolve conflicting reports of both enhancement and quenching of NV$^-$ PL under simultaneous illumination with \SI{532}{\nano\meter} visible and \SI{1064}{\nano\meter} light \cite{Geiselmann2013,Neukirch2013,Lai2013,Ji2016}.  We use this knowledge to demonstrate a new protocol for efficient SCC \textit{via} selective ionization of the NV$^-$ singlet that can lead to high-fidelity, single-shot readout of NV electron spins and drastic reduction of measurement integration time.

Recent attempts to overcome the charge initialization problem include electrical gating \cite{Doi2014} and doping \citep{Doi2016}, but the former yields deterministic initialization only in NV$^0$ and the latter, while effective in stabilizing NV$^-$, also introduces impurities that reduce the spin coherence time. Several schemes have been proposed to address the readout problem. Protocols using quantum logic with nuclear spin ancillae \cite{Jiang2009,Steiner2010} have achieved a 7-fold improvement in SNR over traditional PL measurements \cite{Lovchinsky2016}, at the expense of demanding technical and material requirements. An alternative approach was explored by Shields \textit{et al.} \cite{Shields2015}, who demonstrated spin-to-charge conversion (SCC) through spin-dependent ionization of the NV$^-$ triplet manifold using a visible pulse of light. Together with high-fidelity readout of the NV charge state \cite{Waldherr2011} and extremely high photon collection efficiency, SCC yielded a 4-fold improvement in the single-shot SNR and reduced the spin readout noise to $\sim$3$\times$ the standard quantum limit (SQL). Nonetheless, a room-temperature protocol for single-shot electron-spin readout with SNR$>$1 is still lacking.

\begin{figure}[t]
\includegraphics[scale=1]{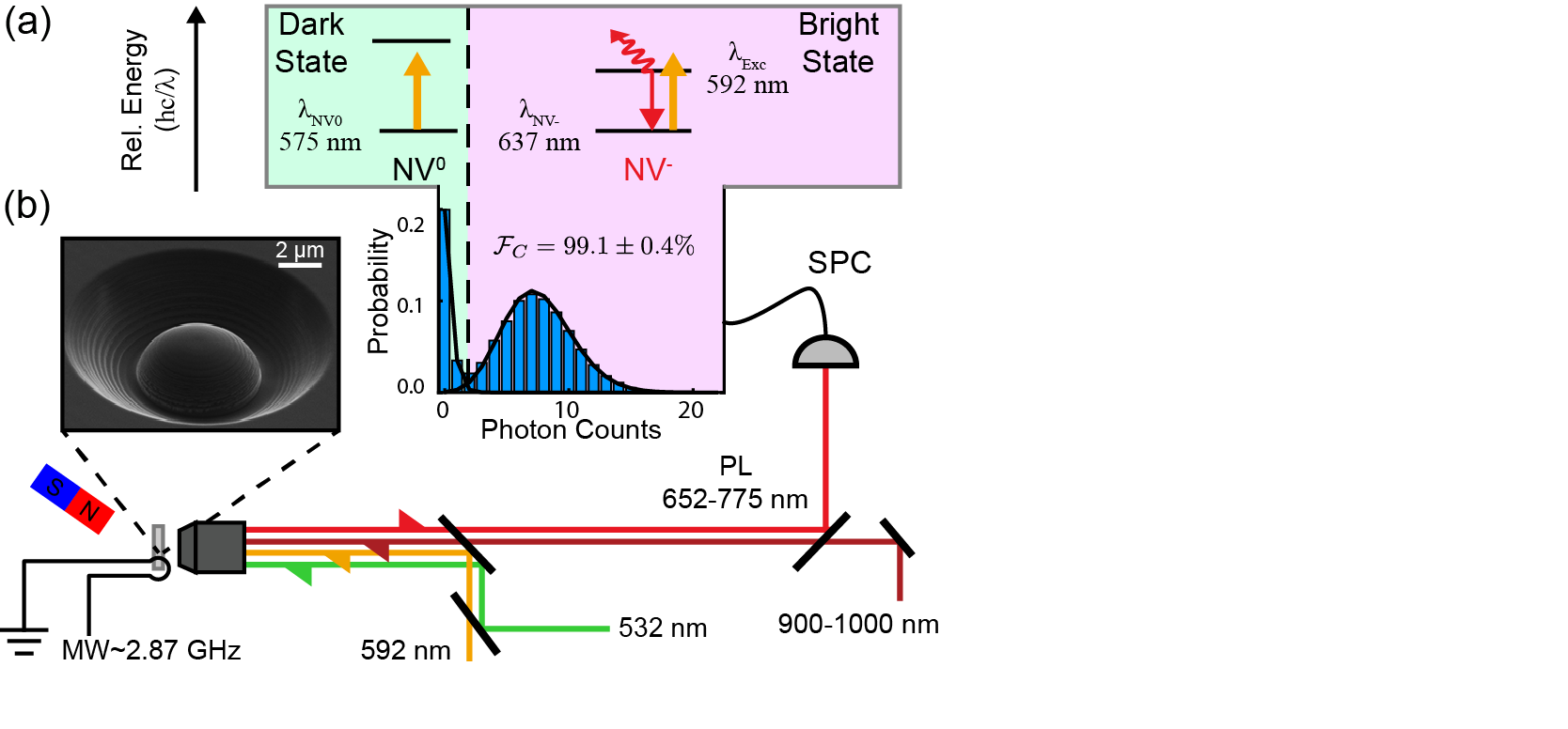}
\caption{Deterministic charge state readout of the NV center. (a) Illumination at \SI{592}{\nano\metre} selectively excites the NV$^-$ charge state (ZPL at \SI{637}{\nano\metre}) and not NV$^0$ (ZPL at \SI{575}{\nano\metre}). A typical photon distribution for a \SI{3}{\milli\second} readout duration is inset below (5000 cycles). (b)  Diagram of the confocal microscope and electron micrograph of the solid immersion lens milled by a focused ion beam from the diamond surface. (ZPL: Zero-phonon line).}
\label{fig:Setup}
\end{figure}



Figure \ref{fig:Setup} illustrates the experimental setup and the concept of high-fidelity charge readout. The technique exploits the blue-shifted absorption spectrum of NV$^0$ compared to NV$^-$, allowing for charge-dependent PL during \SI{592}{\nano\meter} excitation. Precise tuning of the illumination power ($\sim$\SI{100}{\nano\watt}) and readout time produces a strong PL contrast (SNR$\approx$3) that allows for single-shot discrimination of the two charge states by introducing a photon-detection threshold condition (dashed black line in Fig.~\ref{fig:Setup}a). For low illumination power, the single-shot charge measurement is also non-destructive, which facilitates deterministic monitoring of the charge dynamics and the direct measurement of ionization and recombination rates from individual charge-transition events \cite{Aslam2013}. The photon-detection histogram shown in the inset of Fig.~\ref{fig:Setup}a corresponds to a single-shot fidelity $\mathcal{F}_c=99.1\pm0.4\%$ and a non-destructivity exceeding 96\% \cite{Supplemental}.

Individual NVs are addressed using a home-built scanning confocal microscope with three excitation sources (Fig.~\ref{fig:Setup}b). Continuous-wave \SI{532}{\nano\meter} and \SI{592}{\nano\meter} lasers are gated with acousto-optic modulators, and a 900-1000 nm band-pass-filtered supercontinuum source (hereafter termed NIR) produces picosecond pulses with a \SI{40}{\mega\hertz} repetition rate that can be gated in time. A 6-\si{\micro\meter}-diameter solid immersion lens is fabricated around a preselected NV to increase the collection and excitation efficiency using an \textit{in-situ} alignment technique and focused-ion-beam milling. Collected PL is filtered to select for NV$^-$ in the 650-\SI{775}{\nano\meter} band and detected with a single-photon avalanche diode. In this configuration, we record $\sim$0.04 PL photons per \SI{200}{\nano\second} shot \cite{Supplemental}. A $\sim$\SI{20}{\gauss} magnetic field applied along the NV's symmetry axis splits the NV$^-$ ground-state $m_s = \pm1$ spin sublevels, and a \SI{20}{\micro\meter}-diameter gold wire placed across the surface of the diamond is driven by pulsed microwaves to control the ground-state spin.

\begin{figure}[t]
\includegraphics[width=\linewidth]{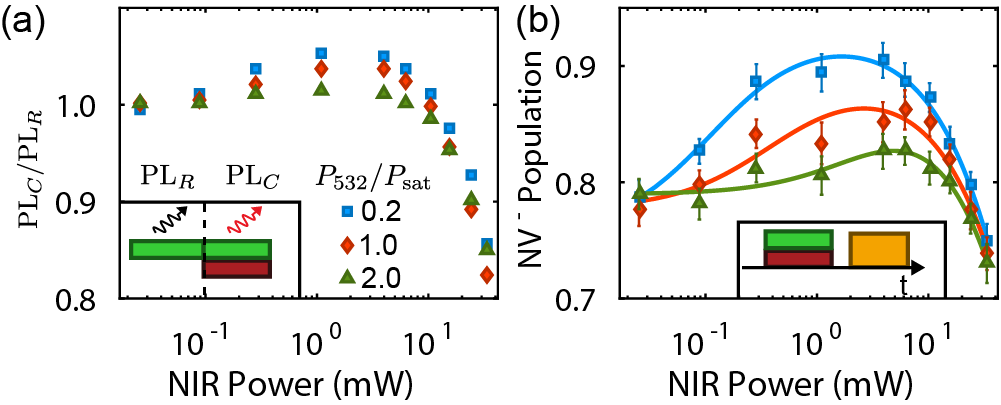}
\caption{Multicolor modulation of steady-state PL and charge. (a) PL due to coincident excitation at \SI{532}{\nano\meter} and 900-1000\ \si{\nano\meter} (PL$_\mathrm{C}$), normalized by the reference PL level under \SI{532}{\nano\meter} excitation alone (PL$_\mathrm{R}$). The NIR excitation is modulated with a square wave at \SI{5}{\kilo\hertz} (inset). Experimental uncertainties are smaller than the symbols. (b) Steady-state population of NV$^-$ (points) for the corresponding power combinations in (a), fit using a model described in the text (curves).}
\label{fig:Quenching}
\end{figure}

Figure~\ref{fig:Quenching} presents measurements of the steady-state PL and corresponding charge distribution under coincident excitation with \SI{532}{\nano\meter} and NIR light. Both sets of measurements exhibit non-monotonic variations as a function of NIR power, P$_{\mathrm{NIR}}$, connecting conflicting observations of PL quenching and enhancement in different regimes \cite{Geiselmann2013,Neukirch2013,Ji2016}.  The direct correlation between the relative changes in PL (Fig.~\ref{fig:Quenching}a) and the underlying charge distributions probed using single-shot readout (Fig.~\ref{fig:Quenching}b) confirm the hypothesized role of charge-state modulation in these effects.  Overall, the non-monotonic response and dependence on green power (P$_{532}$) hint at multiple processes that depend on the relative visible and NIR intensities.  Below, we explore the role of NIR light in modulating the NV's charge dynamics in these different regimes.

Notably, the PL enhancement observed with modest green and NIR powers in Fig.~\ref{fig:Quenching}a is accompanied by an increase in the steady-state NV$^-$ population ($p_\mathrm{minus}$) to a maximum value of 91$\pm0.6\%$, corresponding to an 18\% improvement over the observed population under \SI{532}{\nano\meter} illumination of $77\%$.  This is, to our knowledge, the highest-purity all-optical initialization of NV$^-$ yet reported. Furthermore, we anticipate that the spin purity should be maintained under this protocol since the spin-polarization rate exceeds the charge-switching rate in this regime by over two orders of magnitude \cite{Chen2015a,Supplemental}.

\begin{figure}[t]
\includegraphics[width=\linewidth]{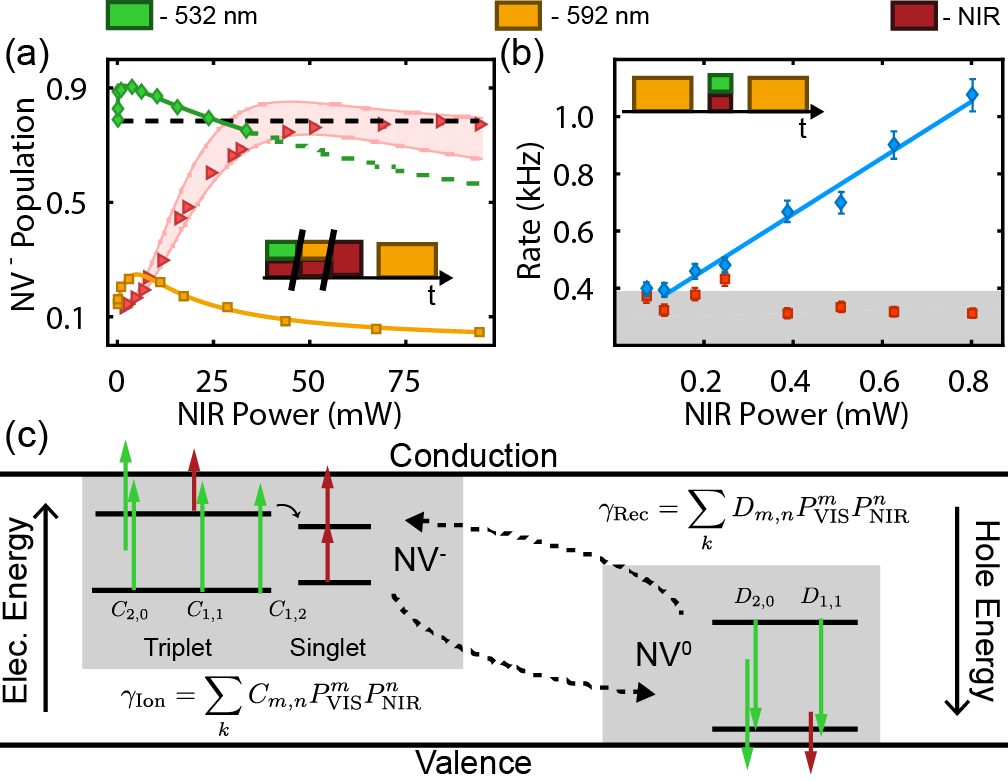}
\caption{Charge dynamics driven by multiphoton absorption. (a) NV$^-$ population versus $P_\mathrm{NIR}$, for fixed visible excitation at \SI{532}{\nano\meter} (\SI{9}{\micro\watt}$\simeq$$0.2 P_\mathrm{sat}$, \protect\includegraphics{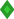}), \SI{592}{\nano\meter} (\SI{20}{\micro\watt}$\simeq \mathrm{0.1} P_\mathrm{sat}$, \protect\includegraphics{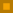}), and for NIR light only (\protect\includegraphics{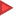}). Curves are fits described in the text. The shaded region indicates the 95\% confidence interval for a simulation of the NIR data using separately measured rates \cite{Supplemental}. (b) Recombination (\protect\includegraphics{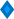}) and ionization (\protect\includegraphics{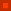}) rates versus $P_\mathrm{NIR}$ for $P_{532}=\SI{5.4}{\micro\watt}\simeq 0.13P_\mathrm{sat}$. The shaded region indicates the noise floor. (c) NV energy diagram indicating allowed optical transitions and corresponding coefficients in the rate model. Insets in (a-b) depict the experimental pulse sequences for each measurement. Except where indicated by error bars, symbol sizes exceed the experimental uncertainty.  }
\label{fig:Enhancement}
\end{figure}

Figure~\ref{fig:Enhancement}a uses a linear horizontal scale to present the same data as Fig.~\ref{fig:Quenching}b (lowest green power) together with an analogous measurement  combining \SI{592}{\nano\meter} and NIR excitation.  We observe an initial enhancement followed by suppression of $p_\mathrm{minus}$ in both cases, which implies this behavior is independent of the vastly different initial conditions produced by \SI{532}{\nano\meter} and \SI{592}{\nano\meter} light alone \cite{Aslam2013}. In contrast, the charge distributions observed in a similar experiment using only NIR light (red triangles in Fig.~\ref{fig:Enhancement}a) exhibit completely different dynamics, underscoring the crucial role of coupled, nonlinear optical processes.

To elucidate the underlying mechanisms, we use high-fidelity, non-destructive charge-state readout to directly measure the NV's ionization and recombination rates.  After non-destructively determining the charge state, we apply an optical pulse with duration and intensity chosen to induce $\ll$1 charge transition, and then measure the resulting state.  Following many such measurements, the rates are calculated from the corresponding transition probabilities divided by the illumination duration.

Figure~\ref{fig:Enhancement}b shows the ionization and recombination rates as a function of P$_{\mathrm{NIR}}$ in the regime of NV$^-$ enhancement ($P_{532}$ is slightly lower than in Fig.~\ref{fig:Enhancement}a to reduce the charge-switching rates due to green light alone, but the steady-state maximum $p_\mathrm{minus}=91\%$ is unchanged).  As expected, the rate for recombination is much larger than for ionization; in fact, the ionization rate is below the noise floor imposed by the $4\%$ destructivity of the charge verification step.

The recombination rate's linear dependence on P$_{\mathrm{NIR}}$ implies that the mechanism driving NV$^-$ enhancement involves a single NIR photon. The most likely candidate is a sequential absorption process in which a \SI{532}{\nano\meter} photon promotes a hole to the excited state of NV$^0$, followed by the absorption of single NIR photon to promote the hole into the valence band, thus converting the center to NV$^-$.  Based on the location of the NV's levels within the bandgap \cite{Aslam2013}, an analogous ionization process is also allowed, but it is apparently $\sim$7 times less likely to occur \cite{Supplemental}.  This asymmetry could result from a combination of the $\sim$50\% longer optical lifetime of NV$^0$ \cite{Liaugaudas2012} and different cross sections for NIR absorption by the NV$^-$ and NV$^0$ excited states. We have also measured the ionization and recombination rates in the presence of NIR light alone \cite{Supplemental}.  They scale with $P_\mathrm{NIR}^3$ and $P_\mathrm{NIR}^2$, respectively, but are several orders of magnitude smaller than the corresponding rates in the presence of visible light.

To capture the effects of these competing nonlinear processes, we employ a master-equation model for the charge-state dynamics, schematically depicted in Fig.~\ref{fig:Enhancement}c.  It includes the allowed orbital and charge transitions considering six levels that comprise the NV$^-$ triplet and singlet manifolds, and the NV$^0$ ground and excited states.
Each ionization or recombination process is assigned a coefficient, $C_{m,n}$ or $D_{m,n}$, respectively, where $m$ and $n$ are the respective number of visible and NIR photons required for that process. For example, the ionization rate for the process corresponding to $C_{m,n}$ is given by $C_{m,n}P^m_{\mathrm{VIS}}P^n_{\mathrm{NIR}}$. The total ionization or recombination rate is the sum of all the individual rates.

We apply this model to fit the steady-state charge distributions in Fig.~\ref{fig:Enhancement}a, using four parameters that quantify the relative weights between the ionization/recombination coefficients \cite{Supplemental}.  Differences between the experiments using \SI{532}{\nano\meter} and \SI{592}{\nano\meter} light are naturally explained by large differences in the cross section for NV$^0$ excitation that affect $D_{1,1}$ and $D_{2,0}$. The charge distributions observed under NIR-only illumination in Fig.~\ref{fig:Enhancement}a do not, in fact, represent the steady-state population, due to the relative weakness of NIR-only nonlinear absorption. Nonetheless, we can quantitatively reproduce those data by adapting our model to account for the slow underlying rates($<$\si{\kilo\hertz}) and a partially destructive charge-state readout \cite{Supplemental}.

Whereas the NV$^-$ enhancement observed at low powers in Fig.~\ref{fig:Enhancement}a is driven by the asymmetry in $D_{1,1}/C_{1,1}$, the suppression at increasing powers results from the higher-order term $C_{1,2}$.  This process corresponds to ionization of NV$^-$ \textit{via} absorption of one visible and two NIR photons. A candidate mechanism is NIR-induced ionization from the metastable singlet ground state.  The singlet manifold is populated through the ISC by visible excitation, and exhibits a zero phonon line at \SI{1042}{\nano\meter} accompanied by a broad phonon-assisted-absorption sideband overlapping our NIR excitation source \cite{Acosta2010,Kehayias2013}.

\begin{figure}
\includegraphics[width=\linewidth]{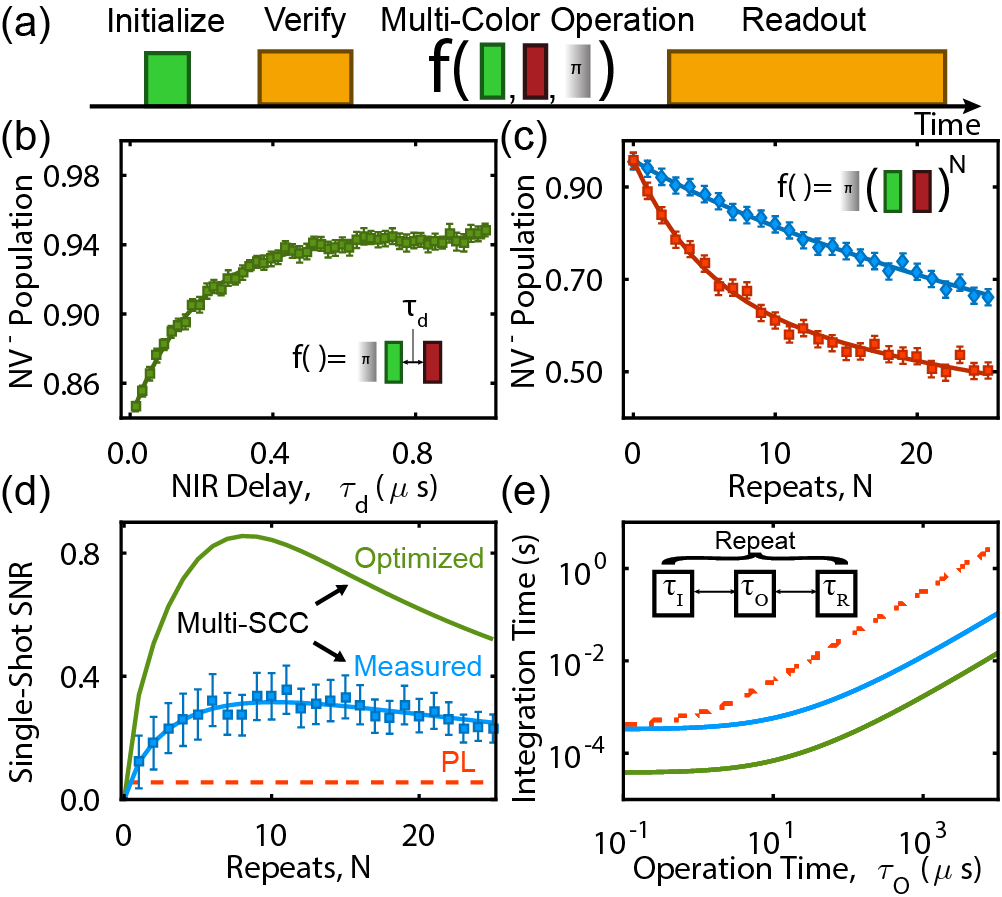}
\caption{Spin-to-charge conversion \textit{via} singlet ionization. (a) Generalized pulse diagram for probing time-domain charge dynamics. (b) Resulting charge state as a function of the delay between a \SI{200}{\nano\second} shelving pulse at \SI{532}{\nano\meter} and a ~\SI{400}{\nano\second} NIR pulse, for a spin prepared in $m_s=-1$. (c) Final charge state as a function of \SI{532}{\nano\meter} shelf duration for spins prepared in $m_s=0$ (\protect\includegraphics[scale=1]{blue_diamond.png}) and $m_s=-1$ (\protect\includegraphics[scale=1]{red_square.png}), and $\tau_d=$ \SI{30}{\nano\second}. (d) Spin-readout SNR comparison of traditional PL (red), measured multi-SCC (blue), and the predicted upper bound for multi-SCC assuming 100\% singlet ionization (green). (e) Required total acquisition time for an SNR=1. The same color legend is used as in (d). The charge state readout duration and power is optimized at each operation time to maximize the speed up.}
\label{fig:SCC}
\end{figure}

To confirm the singlet's role in quenching NV$^-$ at high P$_{\mathrm{NIR}}$, we use the generalized measurement sequence depicted in Fig.~\ref{fig:SCC}a to probe the time-domain ionization dynamics.  Since we are interested in NV$^-$ as a starting state, we use a non-destructive charge-verification step to provide a high-purity, post-selected $p_\mathrm{minus}=96.8\pm0.4\%$. The effects of an arbitrary sequence of visible, NIR, and microwave pulses on the NV's charge are then measured using a high-fidelity readout step.  In Fig.~\ref{fig:SCC}b, we initialize the $m_s=-1$ spin sublevel before populating the singlet with a \SI{200}{\nano\second}, \SI{532}{\nano\meter} shelving pulse separated from a \SI{400}{\nano\second}, \SI{95}{\milli\watt} train of NIR pulses by a variable delay. The resulting $p_\mathrm{minus}$ exhibits exponential decay with a timescale commensurate with the metastable singlet's lifetime ($\tau_{\text{Fit}}=182\pm$\SI{10}{\nano\second}) \cite{Robledo2011,Acosta2010}.

The transient population of the NV$^-$ singlet manifold is highly spin dependent.  Therefore, singlet-selective ionization provides a promising means for SCC.  Indeed, by optimizing the timing of the shelving and ionization pulses, we can achieve a $\sim$7\% spin-dependent contrast in the resulting charge state \cite{Supplemental}.  The contrast is limited in our current setup by the available NIR pulse energy (\SI{2}{\nano\joule}) that ionizes the singlet with per-pulse probability $\sim$6\%. Despite this incomplete ionization, we can enhance the SCC efficiency substantially by repeating the shelve-ionize pulse sequence $N$ times (Fig.~\ref{fig:SCC}c). The spin contrast increases rapidly with $N$ and eventually saturates due to a combination of effects including incomplete ionization, accidental ionization of the triplet, the small ISC probability for $m_s=0$, and imperfect spin initialization.  Fits to the data in Fig.~\ref{fig:SCC}d reflect an extended six-level master-equation model that accounts for all these factors \cite{Supplemental}.

To quantify the performance of multi-SCC spin readout, we consider the single-shot SNR corresponding to a measurement of the spin contrast, i.e., the difference between a spin prepared in $m_s=0$ and $\pm1$ (Fig~\ref{fig:SCC}d). The noise includes contributions from both imperfect SCC efficiency and shot noise in the charge-state readout \cite{Supplemental}.  Our demonstrated protocol exhibits a single-shot $\mathrm{SNR}=0.32$, corresponding to a spin-readout noise 4.6$\times$ the SQL and constituting a 6-fold improvement over traditional PL readout, even in our SIL-enhanced device \cite{Supplemental}.  Given NIR pulses that ionize the singlet with 100\% probability, we predict a further $\sim$2.6-fold improvement to $\mathrm{SNR}=0.83$, corresponding to a single-shot readout fidelity exceeding 75\% and spin-projection noise 1.9$\times$ the SQL. The singlet-selective ionization can be optimized by adjusting the wavelength, pulse width, and repetition rate of the NIR pulse train to compensate the singlet's small optical cross section and short excited-state lifetime ($\sim$\SI{1}{\nano\second}) \cite{Acosta2010}, and with the use of cavities to boost the optical interaction. These values all include the detrimental effect of imperfect spin initialization (we infer $\sim$85\% spin purity). With improved spin purity, multi-SCC should approach a maximum SNR$\sim$1.9, limited by the intrinsic $\sim$10:1 ISC branching ratio.

This dramatic increase in the single-shot SNR comes at the expense of longer readout times. Hence, SCC readout becomes more advantageous for protocols requiring longer spin-operation times between initialization and measurement. We can, however, increase the time-averaged SNR at the expense of the single-shot SNR by reducing the readout time and increasing the \SI{592}{\nano\meter} power, as long as the charge contrast is maintained \cite{Supplemental}. In doing so, we find that even our unoptimized multi-SCC protocol always out-performs traditional PL measurements, in the sense that a shorter integration time is required to achieve the same total SNR (Fig~\ref{fig:SCC}e).  The speedup is especially pronounced for operation times exceeding $\sim$\SI{30}{\micro\second}, where the demonstrated and predicted protocols exhibit $>$10-fold and $>$100-fold improvements over traditional PL, respectively.

In conclusion, we use previously unexplored multi-photon absorption mechanisms to improve both initialization and readout of diamond NV spins. Deliberate tuning of coincident \SI{532}{\nano\meter} and NIR intensities boosts the steady-state NV$^-$ population to $91.0\pm0.6\%$. Carefully timed optical pulse sequences generate efficient spin-to-charge conversion and a universal spin-readout enhancement over the standard approach.  Crucially, these all-optical techniques are applicable to both single-NV and ensemble experiments, and they can lead to significant advances for many research avenues including magnetometry \cite{Lovchinsky2016} and operations involving nuclear spins \cite{Taminiau2012,Kolkowitz2012}, where signal averaging is a critical bottleneck.  As the SCC efficiency approaches the ideal limit of single-shot electron spin readout, it will enable room-temperature applications of protocols that were previously relegated to cryogenic ($<$\SI{10}{\kelvin}) temperatures, including projective and partial measurements of nuclear spins and verification of multi-spin entanglement.

\begin{acknowledgements}
This work was supported by the University of Pennsylvania and a National Science Foundation CAREER grant (EECS-1553511). The authors thank Kameron Oser for helpful discussions.
\end{acknowledgements}

\bibliographystyle{apsrev4-1}
\bibliography{ionization}

\begin{thebibliography}{31}%
\makeatletter
\providecommand \@ifxundefined [1]{%
 \@ifx{#1\undefined}
}%
\providecommand \@ifnum [1]{%
 \ifnum #1\expandafter \@firstoftwo
 \else \expandafter \@secondoftwo
 \fi
}%
\providecommand \@ifx [1]{%
 \ifx #1\expandafter \@firstoftwo
 \else \expandafter \@secondoftwo
 \fi
}%
\providecommand \natexlab [1]{#1}%
\providecommand \enquote  [1]{``#1''}%
\providecommand \bibnamefont  [1]{#1}%
\providecommand \bibfnamefont [1]{#1}%
\providecommand \citenamefont [1]{#1}%
\providecommand \href@noop [0]{\@secondoftwo}%
\providecommand \href [0]{\begingroup \@sanitize@url \@href}%
\providecommand \@href[1]{\@@startlink{#1}\@@href}%
\providecommand \@@href[1]{\endgroup#1\@@endlink}%
\providecommand \@sanitize@url [0]{\catcode `\\12\catcode `\$12\catcode
  `\&12\catcode `\#12\catcode `\^12\catcode `\_12\catcode `\%12\relax}%
\providecommand \@@startlink[1]{}%
\providecommand \@@endlink[0]{}%
\providecommand \url  [0]{\begingroup\@sanitize@url \@url }%
\providecommand \@url [1]{\endgroup\@href {#1}{\urlprefix }}%
\providecommand \urlprefix  [0]{URL }%
\providecommand \Eprint [0]{\href }%
\providecommand \doibase [0]{http://dx.doi.org/}%
\providecommand \selectlanguage [0]{\@gobble}%
\providecommand \bibinfo  [0]{\@secondoftwo}%
\providecommand \bibfield  [0]{\@secondoftwo}%
\providecommand \translation [1]{[#1]}%
\providecommand \BibitemOpen [0]{}%
\providecommand \bibitemStop [0]{}%
\providecommand \bibitemNoStop [0]{.\EOS\space}%
\providecommand \EOS [0]{\spacefactor3000\relax}%
\providecommand \BibitemShut  [1]{\csname bibitem#1\endcsname}%
\let\auto@bib@innerbib\@empty
\bibitem [{\citenamefont {Awschalom}\ \emph {et~al.}(2013)\citenamefont
  {Awschalom}, \citenamefont {Bassett}, \citenamefont {Dzurak}, \citenamefont
  {Hu},\ and\ \citenamefont {Petta}}]{Awschalom2013}%
  \BibitemOpen
  \bibfield  {author} {\bibinfo {author} {\bibfnamefont {D.~D.}\ \bibnamefont
  {Awschalom}}, \bibinfo {author} {\bibfnamefont {L.~C.}\ \bibnamefont
  {Bassett}}, \bibinfo {author} {\bibfnamefont {A.~S.}\ \bibnamefont {Dzurak}},
  \bibinfo {author} {\bibfnamefont {E.~L.}\ \bibnamefont {Hu}}, \ and\ \bibinfo
  {author} {\bibfnamefont {J.~R.}\ \bibnamefont {Petta}},\ }\href {\doibase
  10.1126/science.1231364} {\bibfield  {journal} {\bibinfo  {journal}
  {Science}\ }\textbf {\bibinfo {volume} {339}},\ \bibinfo {pages} {1174}
  (\bibinfo {year} {2013})}\BibitemShut {NoStop}%
\bibitem [{\citenamefont {Lovchinsky}\ \emph {et~al.}(2016)\citenamefont
  {Lovchinsky}, \citenamefont {Sushkov}, \citenamefont {Urbach}, \citenamefont
  {de~Leon}, \citenamefont {Choi}, \citenamefont {De~Greve}, \citenamefont
  {Evans}, \citenamefont {Gertner}, \citenamefont {Bersin}, \citenamefont
  {M{\"u}ller}, \citenamefont {McGuinness}, \citenamefont {Jelezko},
  \citenamefont {Walsworth}, \citenamefont {Park},\ and\ \citenamefont
  {Lukin}}]{Lovchinsky2016}%
  \BibitemOpen
  \bibfield  {author} {\bibinfo {author} {\bibfnamefont {I.}~\bibnamefont
  {Lovchinsky}}, \bibinfo {author} {\bibfnamefont {A.~O.}\ \bibnamefont
  {Sushkov}}, \bibinfo {author} {\bibfnamefont {E.}~\bibnamefont {Urbach}},
  \bibinfo {author} {\bibfnamefont {N.~P.}\ \bibnamefont {de~Leon}}, \bibinfo
  {author} {\bibfnamefont {S.}~\bibnamefont {Choi}}, \bibinfo {author}
  {\bibfnamefont {K.}~\bibnamefont {De~Greve}}, \bibinfo {author}
  {\bibfnamefont {R.}~\bibnamefont {Evans}}, \bibinfo {author} {\bibfnamefont
  {R.}~\bibnamefont {Gertner}}, \bibinfo {author} {\bibfnamefont
  {E.}~\bibnamefont {Bersin}}, \bibinfo {author} {\bibfnamefont
  {C.}~\bibnamefont {M{\"u}ller}}, \bibinfo {author} {\bibfnamefont
  {L.}~\bibnamefont {McGuinness}}, \bibinfo {author} {\bibfnamefont
  {F.}~\bibnamefont {Jelezko}}, \bibinfo {author} {\bibfnamefont {R.~L.}\
  \bibnamefont {Walsworth}}, \bibinfo {author} {\bibfnamefont {H.}~\bibnamefont
  {Park}}, \ and\ \bibinfo {author} {\bibfnamefont {M.~D.}\ \bibnamefont
  {Lukin}},\ }\href {\doibase 10.1126/science.aad8022} {\bibfield  {journal}
  {\bibinfo  {journal} {Science}\ }\textbf {\bibinfo {volume} {351}},\ \bibinfo
  {pages} {836} (\bibinfo {year} {2016})}\BibitemShut {NoStop}%
\bibitem [{\citenamefont {Karaveli}\ \emph {et~al.}(2016)\citenamefont
  {Karaveli}, \citenamefont {Gaathon}, \citenamefont {Wolcott}, \citenamefont
  {Sakakibara}, \citenamefont {Shemesh}, \citenamefont {Peterka}, \citenamefont
  {Boyden}, \citenamefont {Owen}, \citenamefont {Yuste},\ and\ \citenamefont
  {Englund}}]{Karaveli2016}%
  \BibitemOpen
  \bibfield  {author} {\bibinfo {author} {\bibfnamefont {S.}~\bibnamefont
  {Karaveli}}, \bibinfo {author} {\bibfnamefont {O.}~\bibnamefont {Gaathon}},
  \bibinfo {author} {\bibfnamefont {A.}~\bibnamefont {Wolcott}}, \bibinfo
  {author} {\bibfnamefont {R.}~\bibnamefont {Sakakibara}}, \bibinfo {author}
  {\bibfnamefont {O.~A.}\ \bibnamefont {Shemesh}}, \bibinfo {author}
  {\bibfnamefont {D.~S.}\ \bibnamefont {Peterka}}, \bibinfo {author}
  {\bibfnamefont {E.~S.}\ \bibnamefont {Boyden}}, \bibinfo {author}
  {\bibfnamefont {J.~S.}\ \bibnamefont {Owen}}, \bibinfo {author}
  {\bibfnamefont {R.}~\bibnamefont {Yuste}}, \ and\ \bibinfo {author}
  {\bibfnamefont {D.}~\bibnamefont {Englund}},\ }\href {\doibase
  10.1073/pnas.1504451113} {\bibfield  {journal} {\bibinfo  {journal} {Proc.
  Natl. Acad. Sci. USA}\ }\textbf {\bibinfo {volume} {113}},\ \bibinfo {pages}
  {3938} (\bibinfo {year} {2016})}\BibitemShut {NoStop}%
\bibitem [{\citenamefont {DiVincenzo}(2000)}]{DiVincenzo2000}%
  \BibitemOpen
  \bibfield  {author} {\bibinfo {author} {\bibfnamefont {D.~P.}\ \bibnamefont
  {DiVincenzo}},\ }\href {\doibase
  10.1002/1521-3978(200009)48:9/11<771::AID-PROP771>3.0.CO;2-E} {\bibfield
  {journal} {\bibinfo  {journal} {Fortschritte der Physik}\ }\textbf {\bibinfo
  {volume} {48}},\ \bibinfo {pages} {771} (\bibinfo {year} {2000})}\BibitemShut
  {NoStop}%
\bibitem [{\citenamefont {Waldherr}\ \emph
  {et~al.}(2011{\natexlab{a}})\citenamefont {Waldherr}, \citenamefont
  {Neumann}, \citenamefont {Huelga}, \citenamefont {Jelezko},\ and\
  \citenamefont {Wrachtrup}}]{Waldherr2011}%
  \BibitemOpen
  \bibfield  {author} {\bibinfo {author} {\bibfnamefont {G.}~\bibnamefont
  {Waldherr}}, \bibinfo {author} {\bibfnamefont {P.}~\bibnamefont {Neumann}},
  \bibinfo {author} {\bibfnamefont {S.~F.}\ \bibnamefont {Huelga}}, \bibinfo
  {author} {\bibfnamefont {F.}~\bibnamefont {Jelezko}}, \ and\ \bibinfo
  {author} {\bibfnamefont {J.}~\bibnamefont {Wrachtrup}},\ }\href {\doibase
  10.1103/PhysRevLett.107.090401} {\bibfield  {journal} {\bibinfo  {journal}
  {Phys. Rev. Lett.}\ }\textbf {\bibinfo {volume} {107}},\ \bibinfo {pages}
  {090401} (\bibinfo {year} {2011}{\natexlab{a}})}\BibitemShut {NoStop}%
\bibitem [{\citenamefont {Waldherr}\ \emph
  {et~al.}(2011{\natexlab{b}})\citenamefont {Waldherr}, \citenamefont {Beck},
  \citenamefont {Steiner}, \citenamefont {Neumann}, \citenamefont {Gali},
  \citenamefont {Frauenheim}, \citenamefont {Jelezko},\ and\ \citenamefont
  {Wrachtrup}}]{Waldherr2011a}%
  \BibitemOpen
  \bibfield  {author} {\bibinfo {author} {\bibfnamefont {G.}~\bibnamefont
  {Waldherr}}, \bibinfo {author} {\bibfnamefont {J.}~\bibnamefont {Beck}},
  \bibinfo {author} {\bibfnamefont {M.}~\bibnamefont {Steiner}}, \bibinfo
  {author} {\bibfnamefont {P.}~\bibnamefont {Neumann}}, \bibinfo {author}
  {\bibfnamefont {A.}~\bibnamefont {Gali}}, \bibinfo {author} {\bibfnamefont
  {T.}~\bibnamefont {Frauenheim}}, \bibinfo {author} {\bibfnamefont
  {F.}~\bibnamefont {Jelezko}}, \ and\ \bibinfo {author} {\bibfnamefont
  {J.}~\bibnamefont {Wrachtrup}},\ }\href {\doibase
  10.1103/PhysRevLett.106.157601} {\bibfield  {journal} {\bibinfo  {journal}
  {Phys. Rev. Lett.}\ }\textbf {\bibinfo {volume} {106}},\ \bibinfo {pages}
  {157601} (\bibinfo {year} {2011}{\natexlab{b}})}\BibitemShut {NoStop}%
\bibitem [{\citenamefont {Robledo}\ \emph
  {et~al.}(2011{\natexlab{a}})\citenamefont {Robledo}, \citenamefont {Bernien},
  \citenamefont {van~der Sar},\ and\ \citenamefont {Hanson}}]{Robledo2011a}%
  \BibitemOpen
  \bibfield  {author} {\bibinfo {author} {\bibfnamefont {L.}~\bibnamefont
  {Robledo}}, \bibinfo {author} {\bibfnamefont {H.}~\bibnamefont {Bernien}},
  \bibinfo {author} {\bibfnamefont {T.}~\bibnamefont {van~der Sar}}, \ and\
  \bibinfo {author} {\bibfnamefont {R.}~\bibnamefont {Hanson}},\ }\href
  {http://stacks.iop.org/1367-2630/13/i=2/a=025013} {\bibfield  {journal}
  {\bibinfo  {journal} {New J. Phys.}\ }\textbf {\bibinfo {volume} {13}},\
  \bibinfo {pages} {025013} (\bibinfo {year} {2011}{\natexlab{a}})}\BibitemShut
  {NoStop}%
\bibitem [{\citenamefont {Aslam}\ \emph {et~al.}(2013)\citenamefont {Aslam},
  \citenamefont {Waldherr}, \citenamefont {Neumann}, \citenamefont {Jelezko},\
  and\ \citenamefont {Wrachtrup}}]{Aslam2013}%
  \BibitemOpen
  \bibfield  {author} {\bibinfo {author} {\bibfnamefont {N.}~\bibnamefont
  {Aslam}}, \bibinfo {author} {\bibfnamefont {G.}~\bibnamefont {Waldherr}},
  \bibinfo {author} {\bibfnamefont {P.}~\bibnamefont {Neumann}}, \bibinfo
  {author} {\bibfnamefont {F.}~\bibnamefont {Jelezko}}, \ and\ \bibinfo
  {author} {\bibfnamefont {J.}~\bibnamefont {Wrachtrup}},\ }\href
  {http://stacks.iop.org/1367-2630/15/i=1/a=013064} {\bibfield  {journal}
  {\bibinfo  {journal} {New J. Phys.}\ }\textbf {\bibinfo {volume} {15}},\
  \bibinfo {pages} {013064} (\bibinfo {year} {2013})}\BibitemShut {NoStop}%
\bibitem [{\citenamefont {Taylor}\ \emph {et~al.}(2008)\citenamefont {Taylor},
  \citenamefont {Cappellaro}, \citenamefont {Childress}, \citenamefont {Jiang},
  \citenamefont {Budker}, \citenamefont {Hemmer}, \citenamefont {Yacoby},
  \citenamefont {Walsworth},\ and\ \citenamefont {Lukin}}]{Taylor2008}%
  \BibitemOpen
  \bibfield  {author} {\bibinfo {author} {\bibfnamefont {J.~M.}\ \bibnamefont
  {Taylor}}, \bibinfo {author} {\bibfnamefont {P.}~\bibnamefont {Cappellaro}},
  \bibinfo {author} {\bibfnamefont {L.}~\bibnamefont {Childress}}, \bibinfo
  {author} {\bibfnamefont {L.}~\bibnamefont {Jiang}}, \bibinfo {author}
  {\bibfnamefont {D.}~\bibnamefont {Budker}}, \bibinfo {author} {\bibfnamefont
  {P.~R.}\ \bibnamefont {Hemmer}}, \bibinfo {author} {\bibfnamefont
  {A.}~\bibnamefont {Yacoby}}, \bibinfo {author} {\bibfnamefont
  {R.}~\bibnamefont {Walsworth}}, \ and\ \bibinfo {author} {\bibfnamefont
  {M.~D.}\ \bibnamefont {Lukin}},\ }\href {http://dx.doi.org/10.1038/nphys1075}
  {\bibfield  {journal} {\bibinfo  {journal} {Nature Phys.}\ }\textbf {\bibinfo
  {volume} {4}},\ \bibinfo {pages} {810} (\bibinfo {year} {2008})}\BibitemShut
  {NoStop}%
\bibitem [{\citenamefont {Jiang}\ \emph {et~al.}(2009)\citenamefont {Jiang},
  \citenamefont {Hodges}, \citenamefont {Maze}, \citenamefont {Maurer},
  \citenamefont {Taylor}, \citenamefont {Cory}, \citenamefont {Hemmer},
  \citenamefont {Walsworth}, \citenamefont {Yacoby}, \citenamefont {Zibrov},\
  and\ \citenamefont {Lukin}}]{Jiang2009}%
  \BibitemOpen
  \bibfield  {author} {\bibinfo {author} {\bibfnamefont {L.}~\bibnamefont
  {Jiang}}, \bibinfo {author} {\bibfnamefont {J.~S.}\ \bibnamefont {Hodges}},
  \bibinfo {author} {\bibfnamefont {J.~R.}\ \bibnamefont {Maze}}, \bibinfo
  {author} {\bibfnamefont {P.}~\bibnamefont {Maurer}}, \bibinfo {author}
  {\bibfnamefont {J.~M.}\ \bibnamefont {Taylor}}, \bibinfo {author}
  {\bibfnamefont {D.~G.}\ \bibnamefont {Cory}}, \bibinfo {author}
  {\bibfnamefont {P.~R.}\ \bibnamefont {Hemmer}}, \bibinfo {author}
  {\bibfnamefont {R.~L.}\ \bibnamefont {Walsworth}}, \bibinfo {author}
  {\bibfnamefont {A.}~\bibnamefont {Yacoby}}, \bibinfo {author} {\bibfnamefont
  {A.~S.}\ \bibnamefont {Zibrov}}, \ and\ \bibinfo {author} {\bibfnamefont
  {M.~D.}\ \bibnamefont {Lukin}},\ }\href {\doibase 10.1126/science.1176496}
  {\bibfield  {journal} {\bibinfo  {journal} {Science}\ }\textbf {\bibinfo
  {volume} {326}},\ \bibinfo {pages} {267} (\bibinfo {year}
  {2009})}\BibitemShut {NoStop}%
\bibitem [{\citenamefont {van Oort}\ \emph {et~al.}(1988)\citenamefont {van
  Oort}, \citenamefont {Manson},\ and\ \citenamefont {Glasbeek}}]{Oort1988}%
  \BibitemOpen
  \bibfield  {author} {\bibinfo {author} {\bibfnamefont {E.}~\bibnamefont {van
  Oort}}, \bibinfo {author} {\bibfnamefont {N.~B.}\ \bibnamefont {Manson}}, \
  and\ \bibinfo {author} {\bibfnamefont {M.}~\bibnamefont {Glasbeek}},\ }\href
  {http://stacks.iop.org/0022-3719/21/i=23/a=020} {\bibfield  {journal}
  {\bibinfo  {journal} {J. Phys. C: Solid State Phys.}\ }\textbf {\bibinfo
  {volume} {21}},\ \bibinfo {pages} {4385} (\bibinfo {year}
  {1988})}\BibitemShut {NoStop}%
\bibitem [{\citenamefont {Doherty}\ \emph {et~al.}(2013)\citenamefont
  {Doherty}, \citenamefont {Manson}, \citenamefont {Delaney}, \citenamefont
  {Jelezko}, \citenamefont {Wrachtrup},\ and\ \citenamefont
  {Hollenberg}}]{Doherty2013}%
  \BibitemOpen
  \bibfield  {author} {\bibinfo {author} {\bibfnamefont {M.~W.}\ \bibnamefont
  {Doherty}}, \bibinfo {author} {\bibfnamefont {N.~B.}\ \bibnamefont {Manson}},
  \bibinfo {author} {\bibfnamefont {P.}~\bibnamefont {Delaney}}, \bibinfo
  {author} {\bibfnamefont {F.}~\bibnamefont {Jelezko}}, \bibinfo {author}
  {\bibfnamefont {J.}~\bibnamefont {Wrachtrup}}, \ and\ \bibinfo {author}
  {\bibfnamefont {L.~C.}\ \bibnamefont {Hollenberg}},\ }\href {\doibase
  http://dx.doi.org/10.1016/j.physrep.2013.02.001} {\bibfield  {journal}
  {\bibinfo  {journal} {Phys. Rep.}\ }\textbf {\bibinfo {volume} {528}},\
  \bibinfo {pages} {1 } (\bibinfo {year} {2013})}\BibitemShut {NoStop}%
\bibitem [{\citenamefont {Pfaff}\ \emph {et~al.}(2013)\citenamefont {Pfaff},
  \citenamefont {Taminiau}, \citenamefont {Robledo}, \citenamefont {Bernien},
  \citenamefont {Markham}, \citenamefont {Twitchen},\ and\ \citenamefont
  {Hanson}}]{Pfaff2013}%
  \BibitemOpen
  \bibfield  {author} {\bibinfo {author} {\bibfnamefont {W.}~\bibnamefont
  {Pfaff}}, \bibinfo {author} {\bibfnamefont {T.~H.}\ \bibnamefont {Taminiau}},
  \bibinfo {author} {\bibfnamefont {L.}~\bibnamefont {Robledo}}, \bibinfo
  {author} {\bibfnamefont {H.}~\bibnamefont {Bernien}}, \bibinfo {author}
  {\bibfnamefont {M.}~\bibnamefont {Markham}}, \bibinfo {author} {\bibfnamefont
  {D.~J.}\ \bibnamefont {Twitchen}}, \ and\ \bibinfo {author} {\bibfnamefont
  {R.}~\bibnamefont {Hanson}},\ }\href {http://dx.doi.org/10.1038/nphys2444}
  {\bibfield  {journal} {\bibinfo  {journal} {Nature Phys.}\ }\textbf {\bibinfo
  {volume} {9}},\ \bibinfo {pages} {29} (\bibinfo {year} {2013})}\BibitemShut
  {NoStop}%
\bibitem [{\citenamefont {Blok}\ \emph {et~al.}(2014)\citenamefont {Blok},
  \citenamefont {Bonato}, \citenamefont {Markham}, \citenamefont {Twitchen},
  \citenamefont {Dobrovitski},\ and\ \citenamefont {Hanson}}]{Blok2014}%
  \BibitemOpen
  \bibfield  {author} {\bibinfo {author} {\bibfnamefont {M.~S.}\ \bibnamefont
  {Blok}}, \bibinfo {author} {\bibfnamefont {C.}~\bibnamefont {Bonato}},
  \bibinfo {author} {\bibfnamefont {M.~L.}\ \bibnamefont {Markham}}, \bibinfo
  {author} {\bibfnamefont {D.~J.}\ \bibnamefont {Twitchen}}, \bibinfo {author}
  {\bibfnamefont {V.~V.}\ \bibnamefont {Dobrovitski}}, \ and\ \bibinfo {author}
  {\bibfnamefont {R.}~\bibnamefont {Hanson}},\ }\href
  {http://dx.doi.org/10.1038/nphys2881} {\bibfield  {journal} {\bibinfo
  {journal} {Nature Phys.}\ }\textbf {\bibinfo {volume} {10}},\ \bibinfo
  {pages} {189} (\bibinfo {year} {2014})}\BibitemShut {NoStop}%
\bibitem [{\citenamefont {Bernien}\ \emph {et~al.}(2013)\citenamefont
  {Bernien}, \citenamefont {Hensen}, \citenamefont {Pfaff}, \citenamefont
  {Koolstra}, \citenamefont {Blok}, \citenamefont {Robledo}, \citenamefont
  {Taminiau}, \citenamefont {Markham}, \citenamefont {Twitchen}, \citenamefont
  {Childress},\ and\ \citenamefont {Hanson}}]{Bernien2013}%
  \BibitemOpen
  \bibfield  {author} {\bibinfo {author} {\bibfnamefont {H.}~\bibnamefont
  {Bernien}}, \bibinfo {author} {\bibfnamefont {B.}~\bibnamefont {Hensen}},
  \bibinfo {author} {\bibfnamefont {W.}~\bibnamefont {Pfaff}}, \bibinfo
  {author} {\bibfnamefont {G.}~\bibnamefont {Koolstra}}, \bibinfo {author}
  {\bibfnamefont {M.~S.}\ \bibnamefont {Blok}}, \bibinfo {author}
  {\bibfnamefont {L.}~\bibnamefont {Robledo}}, \bibinfo {author} {\bibfnamefont
  {T.~H.}\ \bibnamefont {Taminiau}}, \bibinfo {author} {\bibfnamefont
  {M.}~\bibnamefont {Markham}}, \bibinfo {author} {\bibfnamefont {D.~J.}\
  \bibnamefont {Twitchen}}, \bibinfo {author} {\bibfnamefont {L.}~\bibnamefont
  {Childress}}, \ and\ \bibinfo {author} {\bibfnamefont {R.}~\bibnamefont
  {Hanson}},\ }\href {http://dx.doi.org/10.1038/nature12016} {\bibfield
  {journal} {\bibinfo  {journal} {Nature}\ }\textbf {\bibinfo {volume} {497}},\
  \bibinfo {pages} {86} (\bibinfo {year} {2013})}\BibitemShut {NoStop}%
\bibitem [{\citenamefont {Robledo}\ \emph
  {et~al.}(2011{\natexlab{b}})\citenamefont {Robledo}, \citenamefont
  {Childress}, \citenamefont {Bernien}, \citenamefont {Hensen}, \citenamefont
  {Alkemade},\ and\ \citenamefont {Hanson}}]{Robledo2011}%
  \BibitemOpen
  \bibfield  {author} {\bibinfo {author} {\bibfnamefont {L.}~\bibnamefont
  {Robledo}}, \bibinfo {author} {\bibfnamefont {L.}~\bibnamefont {Childress}},
  \bibinfo {author} {\bibfnamefont {H.}~\bibnamefont {Bernien}}, \bibinfo
  {author} {\bibfnamefont {B.}~\bibnamefont {Hensen}}, \bibinfo {author}
  {\bibfnamefont {P.~F.~A.}\ \bibnamefont {Alkemade}}, \ and\ \bibinfo {author}
  {\bibfnamefont {R.}~\bibnamefont {Hanson}},\ }\href
  {http://dx.doi.org/10.1038/nature10401} {\bibfield  {journal} {\bibinfo
  {journal} {Nature}\ }\textbf {\bibinfo {volume} {477}},\ \bibinfo {pages}
  {574} (\bibinfo {year} {2011}{\natexlab{b}})}\BibitemShut {NoStop}%
\bibitem [{\citenamefont {Geiselmann}\ \emph {et~al.}(2013)\citenamefont
  {Geiselmann}, \citenamefont {Marty}, \citenamefont {Garcia~de Abajo},\ and\
  \citenamefont {Quidant}}]{Geiselmann2013}%
  \BibitemOpen
  \bibfield  {author} {\bibinfo {author} {\bibfnamefont {M.}~\bibnamefont
  {Geiselmann}}, \bibinfo {author} {\bibfnamefont {R.}~\bibnamefont {Marty}},
  \bibinfo {author} {\bibfnamefont {F.~J.}\ \bibnamefont {Garcia~de Abajo}}, \
  and\ \bibinfo {author} {\bibfnamefont {R.}~\bibnamefont {Quidant}},\ }\href
  {http://dx.doi.org/10.1038/nphys2770} {\bibfield  {journal} {\bibinfo
  {journal} {Nature Phys.}\ }\textbf {\bibinfo {volume} {9}},\ \bibinfo {pages}
  {785} (\bibinfo {year} {2013})}\BibitemShut {NoStop}%
\bibitem [{\citenamefont {Neukirch}\ \emph {et~al.}(2013)\citenamefont
  {Neukirch}, \citenamefont {Gieseler}, \citenamefont {Quidant}, \citenamefont
  {Novotny},\ and\ \citenamefont {Vamivakas}}]{Neukirch2013}%
  \BibitemOpen
  \bibfield  {author} {\bibinfo {author} {\bibfnamefont {L.~P.}\ \bibnamefont
  {Neukirch}}, \bibinfo {author} {\bibfnamefont {J.}~\bibnamefont {Gieseler}},
  \bibinfo {author} {\bibfnamefont {R.}~\bibnamefont {Quidant}}, \bibinfo
  {author} {\bibfnamefont {L.}~\bibnamefont {Novotny}}, \ and\ \bibinfo
  {author} {\bibfnamefont {A.~N.}\ \bibnamefont {Vamivakas}},\ }\href {\doibase
  10.1364/OL.38.002976} {\bibfield  {journal} {\bibinfo  {journal} {Opt.
  Lett.}\ }\textbf {\bibinfo {volume} {38}},\ \bibinfo {pages} {2976} (\bibinfo
  {year} {2013})}\BibitemShut {NoStop}%
\bibitem [{\citenamefont {Lai}\ \emph {et~al.}(2013)\citenamefont {Lai},
  \citenamefont {Faklaris}, \citenamefont {Zheng}, \citenamefont {Jacques},
  \citenamefont {Chang}, \citenamefont {Roch},\ and\ \citenamefont
  {Treussart}}]{Lai2013}%
  \BibitemOpen
  \bibfield  {author} {\bibinfo {author} {\bibfnamefont {N.~D.}\ \bibnamefont
  {Lai}}, \bibinfo {author} {\bibfnamefont {O.}~\bibnamefont {Faklaris}},
  \bibinfo {author} {\bibfnamefont {D.}~\bibnamefont {Zheng}}, \bibinfo
  {author} {\bibfnamefont {V.}~\bibnamefont {Jacques}}, \bibinfo {author}
  {\bibfnamefont {H.-C.}\ \bibnamefont {Chang}}, \bibinfo {author}
  {\bibfnamefont {J.-F.}\ \bibnamefont {Roch}}, \ and\ \bibinfo {author}
  {\bibfnamefont {F.}~\bibnamefont {Treussart}},\ }\href
  {http://stacks.iop.org/1367-2630/15/i=3/a=033030} {\bibfield  {journal}
  {\bibinfo  {journal} {New Journal of Physics}\ }\textbf {\bibinfo {volume}
  {15}},\ \bibinfo {pages} {033030} (\bibinfo {year} {2013})}\BibitemShut
  {NoStop}%
\bibitem [{\citenamefont {Ji}\ and\ \citenamefont {Dutt}(2016)}]{Ji2016}%
  \BibitemOpen
  \bibfield  {author} {\bibinfo {author} {\bibfnamefont {P.}~\bibnamefont
  {Ji}}\ and\ \bibinfo {author} {\bibfnamefont {M.~V.~G.}\ \bibnamefont
  {Dutt}},\ }\href {\doibase 10.1103/PhysRevB.94.024101} {\bibfield  {journal}
  {\bibinfo  {journal} {Phys. Rev. B}\ }\textbf {\bibinfo {volume} {94}},\
  \bibinfo {pages} {024101} (\bibinfo {year} {2016})}\BibitemShut {NoStop}%
\bibitem [{\citenamefont {Doi}\ \emph {et~al.}(2014)\citenamefont {Doi},
  \citenamefont {Makino}, \citenamefont {Kato}, \citenamefont {Takeuchi},
  \citenamefont {Ogura}, \citenamefont {Okushi}, \citenamefont {Morishita},
  \citenamefont {Tashima}, \citenamefont {Miwa}, \citenamefont {Yamasaki},
  \citenamefont {Neumann}, \citenamefont {Wrachtrup}, \citenamefont {Suzuki},\
  and\ \citenamefont {Mizuochi}}]{Doi2014}%
  \BibitemOpen
  \bibfield  {author} {\bibinfo {author} {\bibfnamefont {Y.}~\bibnamefont
  {Doi}}, \bibinfo {author} {\bibfnamefont {T.}~\bibnamefont {Makino}},
  \bibinfo {author} {\bibfnamefont {H.}~\bibnamefont {Kato}}, \bibinfo {author}
  {\bibfnamefont {D.}~\bibnamefont {Takeuchi}}, \bibinfo {author}
  {\bibfnamefont {M.}~\bibnamefont {Ogura}}, \bibinfo {author} {\bibfnamefont
  {H.}~\bibnamefont {Okushi}}, \bibinfo {author} {\bibfnamefont
  {H.}~\bibnamefont {Morishita}}, \bibinfo {author} {\bibfnamefont
  {T.}~\bibnamefont {Tashima}}, \bibinfo {author} {\bibfnamefont
  {S.}~\bibnamefont {Miwa}}, \bibinfo {author} {\bibfnamefont {S.}~\bibnamefont
  {Yamasaki}}, \bibinfo {author} {\bibfnamefont {P.}~\bibnamefont {Neumann}},
  \bibinfo {author} {\bibfnamefont {J.}~\bibnamefont {Wrachtrup}}, \bibinfo
  {author} {\bibfnamefont {Y.}~\bibnamefont {Suzuki}}, \ and\ \bibinfo {author}
  {\bibfnamefont {N.}~\bibnamefont {Mizuochi}},\ }\href {\doibase
  10.1103/PhysRevX.4.011057} {\bibfield  {journal} {\bibinfo  {journal} {Phys.
  Rev. X}\ }\textbf {\bibinfo {volume} {4}},\ \bibinfo {pages} {011057}
  (\bibinfo {year} {2014})}\BibitemShut {NoStop}%
\bibitem [{\citenamefont {Doi}\ \emph {et~al.}(2016)\citenamefont {Doi},
  \citenamefont {Fukui}, \citenamefont {Kato}, \citenamefont {Makino},
  \citenamefont {Yamasaki}, \citenamefont {Tashima}, \citenamefont {Morishita},
  \citenamefont {Miwa}, \citenamefont {Jelezko}, \citenamefont {Suzuki},\ and\
  \citenamefont {Mizuochi}}]{Doi2016}%
  \BibitemOpen
  \bibfield  {author} {\bibinfo {author} {\bibfnamefont {Y.}~\bibnamefont
  {Doi}}, \bibinfo {author} {\bibfnamefont {T.}~\bibnamefont {Fukui}}, \bibinfo
  {author} {\bibfnamefont {H.}~\bibnamefont {Kato}}, \bibinfo {author}
  {\bibfnamefont {T.}~\bibnamefont {Makino}}, \bibinfo {author} {\bibfnamefont
  {S.}~\bibnamefont {Yamasaki}}, \bibinfo {author} {\bibfnamefont
  {T.}~\bibnamefont {Tashima}}, \bibinfo {author} {\bibfnamefont
  {H.}~\bibnamefont {Morishita}}, \bibinfo {author} {\bibfnamefont
  {S.}~\bibnamefont {Miwa}}, \bibinfo {author} {\bibfnamefont {F.}~\bibnamefont
  {Jelezko}}, \bibinfo {author} {\bibfnamefont {Y.}~\bibnamefont {Suzuki}}, \
  and\ \bibinfo {author} {\bibfnamefont {N.}~\bibnamefont {Mizuochi}},\ }\href
  {\doibase 10.1103/PhysRevB.93.081203} {\bibfield  {journal} {\bibinfo
  {journal} {Phys. Rev. B}\ }\textbf {\bibinfo {volume} {93}},\ \bibinfo
  {pages} {081203} (\bibinfo {year} {2016})}\BibitemShut {NoStop}%
\bibitem [{\citenamefont {Steiner}\ \emph {et~al.}(2010)\citenamefont
  {Steiner}, \citenamefont {Neumann}, \citenamefont {Beck}, \citenamefont
  {Jelezko},\ and\ \citenamefont {Wrachtrup}}]{Steiner2010}%
  \BibitemOpen
  \bibfield  {author} {\bibinfo {author} {\bibfnamefont {M.}~\bibnamefont
  {Steiner}}, \bibinfo {author} {\bibfnamefont {P.}~\bibnamefont {Neumann}},
  \bibinfo {author} {\bibfnamefont {J.}~\bibnamefont {Beck}}, \bibinfo {author}
  {\bibfnamefont {F.}~\bibnamefont {Jelezko}}, \ and\ \bibinfo {author}
  {\bibfnamefont {J.}~\bibnamefont {Wrachtrup}},\ }\href {\doibase
  10.1103/PhysRevB.81.035205} {\bibfield  {journal} {\bibinfo  {journal} {Phys.
  Rev. B}\ }\textbf {\bibinfo {volume} {81}},\ \bibinfo {pages} {035205}
  (\bibinfo {year} {2010})}\BibitemShut {NoStop}%
\bibitem [{\citenamefont {Shields}\ \emph {et~al.}(2015)\citenamefont
  {Shields}, \citenamefont {Unterreithmeier}, \citenamefont {de~Leon},
  \citenamefont {Park},\ and\ \citenamefont {Lukin}}]{Shields2015}%
  \BibitemOpen
  \bibfield  {author} {\bibinfo {author} {\bibfnamefont {B.~J.}\ \bibnamefont
  {Shields}}, \bibinfo {author} {\bibfnamefont {Q.~P.}\ \bibnamefont
  {Unterreithmeier}}, \bibinfo {author} {\bibfnamefont {N.~P.}\ \bibnamefont
  {de~Leon}}, \bibinfo {author} {\bibfnamefont {H.}~\bibnamefont {Park}}, \
  and\ \bibinfo {author} {\bibfnamefont {M.~D.}\ \bibnamefont {Lukin}},\ }\href
  {http://journals.aps.org/prl/abstract/10.1103/PhysRevLett.114.136402}
  {\bibfield  {journal} {\bibinfo  {journal} {Phys. Rev. Lett.}\ }\textbf
  {\bibinfo {volume} {114}},\ \bibinfo {pages} {136402} (\bibinfo {year}
  {2015})}\BibitemShut {NoStop}%
\bibitem [{Sup()}]{Supplemental}%
  \BibitemOpen
  \href@noop {} {\ }\bibinfo {note} {See the supplemental information online
  for further details.}\BibitemShut {Stop}%
\bibitem [{\citenamefont {Chen}\ \emph {et~al.}(2015)\citenamefont {Chen},
  \citenamefont {Zhou}, \citenamefont {Zou}, \citenamefont {Li}, \citenamefont
  {Dong}, \citenamefont {Sun},\ and\ \citenamefont {Guo}}]{Chen2015a}%
  \BibitemOpen
  \bibfield  {author} {\bibinfo {author} {\bibfnamefont {X.-D.}\ \bibnamefont
  {Chen}}, \bibinfo {author} {\bibfnamefont {L.-M.}\ \bibnamefont {Zhou}},
  \bibinfo {author} {\bibfnamefont {C.-L.}\ \bibnamefont {Zou}}, \bibinfo
  {author} {\bibfnamefont {C.-C.}\ \bibnamefont {Li}}, \bibinfo {author}
  {\bibfnamefont {Y.}~\bibnamefont {Dong}}, \bibinfo {author} {\bibfnamefont
  {F.-W.}\ \bibnamefont {Sun}}, \ and\ \bibinfo {author} {\bibfnamefont
  {G.-C.}\ \bibnamefont {Guo}},\ }\href {\doibase 10.1103/PhysRevB.92.104301}
  {\bibfield  {journal} {\bibinfo  {journal} {Phys. Rev. B}\ }\textbf {\bibinfo
  {volume} {92}},\ \bibinfo {pages} {104301} (\bibinfo {year}
  {2015})}\BibitemShut {NoStop}%
\bibitem [{\citenamefont {Liaugaudas}\ \emph {et~al.}(2012)\citenamefont
  {Liaugaudas}, \citenamefont {Davies}, \citenamefont {Suhling}, \citenamefont
  {Khan},\ and\ \citenamefont {Evans}}]{Liaugaudas2012}%
  \BibitemOpen
  \bibfield  {author} {\bibinfo {author} {\bibfnamefont {G.}~\bibnamefont
  {Liaugaudas}}, \bibinfo {author} {\bibfnamefont {G.}~\bibnamefont {Davies}},
  \bibinfo {author} {\bibfnamefont {K.}~\bibnamefont {Suhling}}, \bibinfo
  {author} {\bibfnamefont {R.~U.~A.}\ \bibnamefont {Khan}}, \ and\ \bibinfo
  {author} {\bibfnamefont {D.~J.~F.}\ \bibnamefont {Evans}},\ }\href
  {http://stacks.iop.org/0953-8984/24/i=43/a=435503} {\bibfield  {journal}
  {\bibinfo  {journal} {J.Phys.: Condens. Matter}\ }\textbf {\bibinfo {volume}
  {24}},\ \bibinfo {pages} {435503} (\bibinfo {year} {2012})}\BibitemShut
  {NoStop}%
\bibitem [{\citenamefont {Acosta}\ \emph {et~al.}(2010)\citenamefont {Acosta},
  \citenamefont {Jarmola}, \citenamefont {Bauch},\ and\ \citenamefont
  {Budker}}]{Acosta2010}%
  \BibitemOpen
  \bibfield  {author} {\bibinfo {author} {\bibfnamefont {V.~M.}\ \bibnamefont
  {Acosta}}, \bibinfo {author} {\bibfnamefont {A.}~\bibnamefont {Jarmola}},
  \bibinfo {author} {\bibfnamefont {E.}~\bibnamefont {Bauch}}, \ and\ \bibinfo
  {author} {\bibfnamefont {D.}~\bibnamefont {Budker}},\ }\href {\doibase
  10.1103/PhysRevB.82.201202} {\bibfield  {journal} {\bibinfo  {journal} {Phys.
  Rev. B}\ }\textbf {\bibinfo {volume} {82}},\ \bibinfo {pages} {201202}
  (\bibinfo {year} {2010})}\BibitemShut {NoStop}%
\bibitem [{\citenamefont {Kehayias}\ \emph {et~al.}(2013)\citenamefont
  {Kehayias}, \citenamefont {Doherty}, \citenamefont {English}, \citenamefont
  {Fischer}, \citenamefont {Jarmola}, \citenamefont {Jensen}, \citenamefont
  {Leefer}, \citenamefont {Hemmer}, \citenamefont {Manson},\ and\ \citenamefont
  {Budker}}]{Kehayias2013}%
  \BibitemOpen
  \bibfield  {author} {\bibinfo {author} {\bibfnamefont {P.}~\bibnamefont
  {Kehayias}}, \bibinfo {author} {\bibfnamefont {M.~W.}\ \bibnamefont
  {Doherty}}, \bibinfo {author} {\bibfnamefont {D.}~\bibnamefont {English}},
  \bibinfo {author} {\bibfnamefont {R.}~\bibnamefont {Fischer}}, \bibinfo
  {author} {\bibfnamefont {A.}~\bibnamefont {Jarmola}}, \bibinfo {author}
  {\bibfnamefont {K.}~\bibnamefont {Jensen}}, \bibinfo {author} {\bibfnamefont
  {N.}~\bibnamefont {Leefer}}, \bibinfo {author} {\bibfnamefont
  {P.}~\bibnamefont {Hemmer}}, \bibinfo {author} {\bibfnamefont {N.~B.}\
  \bibnamefont {Manson}}, \ and\ \bibinfo {author} {\bibfnamefont
  {D.}~\bibnamefont {Budker}},\ }\href
  {http://journals.aps.org/prb/abstract/10.1103/PhysRevB.88.165202} {\bibfield
  {journal} {\bibinfo  {journal} {Phys. Rev. B}\ }\textbf {\bibinfo {volume}
  {88}},\ \bibinfo {pages} {165202} (\bibinfo {year} {2013})}\BibitemShut
  {NoStop}%
\bibitem [{\citenamefont {Taminiau}\ \emph {et~al.}(2012)\citenamefont
  {Taminiau}, \citenamefont {Wagenaar}, \citenamefont {van~der Sar},
  \citenamefont {Jelezko}, \citenamefont {Dobrovitski},\ and\ \citenamefont
  {Hanson}}]{Taminiau2012}%
  \BibitemOpen
  \bibfield  {author} {\bibinfo {author} {\bibfnamefont {T.~H.}\ \bibnamefont
  {Taminiau}}, \bibinfo {author} {\bibfnamefont {J.~J.~T.}\ \bibnamefont
  {Wagenaar}}, \bibinfo {author} {\bibfnamefont {T.}~\bibnamefont {van~der
  Sar}}, \bibinfo {author} {\bibfnamefont {F.}~\bibnamefont {Jelezko}},
  \bibinfo {author} {\bibfnamefont {V.~V.}\ \bibnamefont {Dobrovitski}}, \ and\
  \bibinfo {author} {\bibfnamefont {R.}~\bibnamefont {Hanson}},\ }\href
  {\doibase 10.1103/PhysRevLett.109.137602} {\bibfield  {journal} {\bibinfo
  {journal} {Phys. Rev. Lett.}\ }\textbf {\bibinfo {volume} {109}},\ \bibinfo
  {pages} {137602} (\bibinfo {year} {2012})}\BibitemShut {NoStop}%
\bibitem [{\citenamefont {Kolkowitz}\ \emph {et~al.}(2012)\citenamefont
  {Kolkowitz}, \citenamefont {Unterreithmeier}, \citenamefont {Bennett},\ and\
  \citenamefont {Lukin}}]{Kolkowitz2012}%
  \BibitemOpen
  \bibfield  {author} {\bibinfo {author} {\bibfnamefont {S.}~\bibnamefont
  {Kolkowitz}}, \bibinfo {author} {\bibfnamefont {Q.~P.}\ \bibnamefont
  {Unterreithmeier}}, \bibinfo {author} {\bibfnamefont {S.~D.}\ \bibnamefont
  {Bennett}}, \ and\ \bibinfo {author} {\bibfnamefont {M.~D.}\ \bibnamefont
  {Lukin}},\ }\href {\doibase 10.1103/PhysRevLett.109.137601} {\bibfield
  {journal} {\bibinfo  {journal} {Phys. Rev. Lett.}\ }\textbf {\bibinfo
  {volume} {109}},\ \bibinfo {pages} {137601} (\bibinfo {year}
  {2012})}\BibitemShut {NoStop}%
\end{thebibliography}%


\begin{thebibliography}{8}%
\makeatletter
\providecommand \@ifxundefined [1]{%
 \@ifx{#1\undefined}
}%
\providecommand \@ifnum [1]{%
 \ifnum #1\expandafter \@firstoftwo
 \else \expandafter \@secondoftwo
 \fi
}%
\providecommand \@ifx [1]{%
 \ifx #1\expandafter \@firstoftwo
 \else \expandafter \@secondoftwo
 \fi
}%
\providecommand \natexlab [1]{#1}%
\providecommand \enquote  [1]{``#1''}%
\providecommand \bibnamefont  [1]{#1}%
\providecommand \bibfnamefont [1]{#1}%
\providecommand \citenamefont [1]{#1}%
\providecommand \href@noop [0]{\@secondoftwo}%
\providecommand \href [0]{\begingroup \@sanitize@url \@href}%
\providecommand \@href[1]{\@@startlink{#1}\@@href}%
\providecommand \@@href[1]{\endgroup#1\@@endlink}%
\providecommand \@sanitize@url [0]{\catcode `\\12\catcode `\$12\catcode
  `\&12\catcode `\#12\catcode `\^12\catcode `\_12\catcode `\%12\relax}%
\providecommand \@@startlink[1]{}%
\providecommand \@@endlink[0]{}%
\providecommand \url  [0]{\begingroup\@sanitize@url \@url }%
\providecommand \@url [1]{\endgroup\@href {#1}{\urlprefix }}%
\providecommand \urlprefix  [0]{URL }%
\providecommand \Eprint [0]{\href }%
\providecommand \doibase [0]{http://dx.doi.org/}%
\providecommand \selectlanguage [0]{\@gobble}%
\providecommand \bibinfo  [0]{\@secondoftwo}%
\providecommand \bibfield  [0]{\@secondoftwo}%
\providecommand \translation [1]{[#1]}%
\providecommand \BibitemOpen [0]{}%
\providecommand \bibitemStop [0]{}%
\providecommand \bibitemNoStop [0]{.\EOS\space}%
\providecommand \EOS [0]{\spacefactor3000\relax}%
\providecommand \BibitemShut  [1]{\csname bibitem#1\endcsname}%
\let\auto@bib@innerbib\@empty
\bibitem [{\citenamefont {Jamali}\ \emph {et~al.}(2014)\citenamefont {Jamali},
  \citenamefont {Gerhardt}, \citenamefont {Rezai}, \citenamefont {Frenner},
  \citenamefont {Fedder},\ and\ \citenamefont {Wrachtrup}}]{Jamali2014}%
  \BibitemOpen
  \bibfield  {author} {\bibinfo {author} {\bibfnamefont {M.}~\bibnamefont
  {Jamali}}, \bibinfo {author} {\bibfnamefont {I.}~\bibnamefont {Gerhardt}},
  \bibinfo {author} {\bibfnamefont {M.}~\bibnamefont {Rezai}}, \bibinfo
  {author} {\bibfnamefont {K.}~\bibnamefont {Frenner}}, \bibinfo {author}
  {\bibfnamefont {H.}~\bibnamefont {Fedder}}, \ and\ \bibinfo {author}
  {\bibfnamefont {J.}~\bibnamefont {Wrachtrup}},\ }\href
  {http://scitation.aip.org/content/aip/journal/rsi/85/12/10.1063/1.4902818}
  {\bibfield  {journal} {\bibinfo  {journal} {Rev. Sci. Instrum.}\ }\textbf
  {\bibinfo {volume} {85}},\ \bibinfo {eid} {123703} (\bibinfo {year}
  {2014})}\BibitemShut {NoStop}%
\bibitem [{\citenamefont {Lee}\ \emph {et~al.}(2008)\citenamefont {Lee},
  \citenamefont {Gu}, \citenamefont {Dawson}, \citenamefont {Friel},\ and\
  \citenamefont {Scarsbrook}}]{Lee2008}%
  \BibitemOpen
  \bibfield  {author} {\bibinfo {author} {\bibfnamefont {C.}~\bibnamefont
  {Lee}}, \bibinfo {author} {\bibfnamefont {E.}~\bibnamefont {Gu}}, \bibinfo
  {author} {\bibfnamefont {M.}~\bibnamefont {Dawson}}, \bibinfo {author}
  {\bibfnamefont {I.}~\bibnamefont {Friel}}, \ and\ \bibinfo {author}
  {\bibfnamefont {G.}~\bibnamefont {Scarsbrook}},\ }\href {\doibase
  http://dx.doi.org/10.1016/j.diamond.2008.01.011} {\bibfield  {journal}
  {\bibinfo  {journal} {Diamond Relat. Mater.}\ }\textbf {\bibinfo {volume}
  {17}},\ \bibinfo {pages} {1292 } (\bibinfo {year} {2008})}\BibitemShut
  {NoStop}%
\bibitem [{\citenamefont {Ivanov}\ \emph {et~al.}(2013)\citenamefont {Ivanov},
  \citenamefont {Li}, \citenamefont {Dolan},\ and\ \citenamefont
  {Gu}}]{Ivanov2013}%
  \BibitemOpen
  \bibfield  {author} {\bibinfo {author} {\bibfnamefont {I.}~\bibnamefont
  {Ivanov}}, \bibinfo {author} {\bibfnamefont {X.}~\bibnamefont {Li}}, \bibinfo
  {author} {\bibfnamefont {P.}~\bibnamefont {Dolan}}, \ and\ \bibinfo {author}
  {\bibfnamefont {M.}~\bibnamefont {Gu}},\ }\href
  {https://www.osapublishing.org/ol/abstract.cfm?uri=ol-38-8-1358} {\bibfield
  {journal} {\bibinfo  {journal} {Opt. Lett.}\ }\textbf {\bibinfo {volume}
  {34}},\ \bibinfo {pages} {1358} (\bibinfo {year} {2013})}\BibitemShut
  {NoStop}%
\bibitem [{\citenamefont {Aslam}\ \emph {et~al.}(2013)\citenamefont {Aslam},
  \citenamefont {Waldherr}, \citenamefont {Neumann}, \citenamefont {Jelezko},\
  and\ \citenamefont {Wrachtrup}}]{Aslam2013}%
  \BibitemOpen
  \bibfield  {author} {\bibinfo {author} {\bibfnamefont {N.}~\bibnamefont
  {Aslam}}, \bibinfo {author} {\bibfnamefont {G.}~\bibnamefont {Waldherr}},
  \bibinfo {author} {\bibfnamefont {P.}~\bibnamefont {Neumann}}, \bibinfo
  {author} {\bibfnamefont {F.}~\bibnamefont {Jelezko}}, \ and\ \bibinfo
  {author} {\bibfnamefont {J.}~\bibnamefont {Wrachtrup}},\ }\href
  {http://stacks.iop.org/1367-2630/15/i=1/a=013064} {\bibfield  {journal}
  {\bibinfo  {journal} {New J. Phys.}\ }\textbf {\bibinfo {volume} {15}},\
  \bibinfo {pages} {013064} (\bibinfo {year} {2013})}\BibitemShut {NoStop}%
\bibitem [{\citenamefont {Shields}\ \emph {et~al.}(2015)\citenamefont
  {Shields}, \citenamefont {Unterreithmeier}, \citenamefont {de~Leon},
  \citenamefont {Park},\ and\ \citenamefont {Lukin}}]{Shields2015}%
  \BibitemOpen
  \bibfield  {author} {\bibinfo {author} {\bibfnamefont {B.~J.}\ \bibnamefont
  {Shields}}, \bibinfo {author} {\bibfnamefont {Q.~P.}\ \bibnamefont
  {Unterreithmeier}}, \bibinfo {author} {\bibfnamefont {N.~P.}\ \bibnamefont
  {de~Leon}}, \bibinfo {author} {\bibfnamefont {H.}~\bibnamefont {Park}}, \
  and\ \bibinfo {author} {\bibfnamefont {M.~D.}\ \bibnamefont {Lukin}},\ }\href
  {http://journals.aps.org/prl/abstract/10.1103/PhysRevLett.114.136402}
  {\bibfield  {journal} {\bibinfo  {journal} {Phys. Rev. Lett.}\ }\textbf
  {\bibinfo {volume} {114}},\ \bibinfo {pages} {136402} (\bibinfo {year}
  {2015})}\BibitemShut {NoStop}%
\bibitem [{\citenamefont {Lovchinsky}\ \emph {et~al.}(2016)\citenamefont
  {Lovchinsky}, \citenamefont {Sushkov}, \citenamefont {Urbach}, \citenamefont
  {de~Leon}, \citenamefont {Choi}, \citenamefont {De~Greve}, \citenamefont
  {Evans}, \citenamefont {Gertner}, \citenamefont {Bersin}, \citenamefont
  {M{\"u}ller}, \citenamefont {McGuinness}, \citenamefont {Jelezko},
  \citenamefont {Walsworth}, \citenamefont {Park},\ and\ \citenamefont
  {Lukin}}]{Lovchinsky2016}%
  \BibitemOpen
  \bibfield  {author} {\bibinfo {author} {\bibfnamefont {I.}~\bibnamefont
  {Lovchinsky}}, \bibinfo {author} {\bibfnamefont {A.~O.}\ \bibnamefont
  {Sushkov}}, \bibinfo {author} {\bibfnamefont {E.}~\bibnamefont {Urbach}},
  \bibinfo {author} {\bibfnamefont {N.~P.}\ \bibnamefont {de~Leon}}, \bibinfo
  {author} {\bibfnamefont {S.}~\bibnamefont {Choi}}, \bibinfo {author}
  {\bibfnamefont {K.}~\bibnamefont {De~Greve}}, \bibinfo {author}
  {\bibfnamefont {R.}~\bibnamefont {Evans}}, \bibinfo {author} {\bibfnamefont
  {R.}~\bibnamefont {Gertner}}, \bibinfo {author} {\bibfnamefont
  {E.}~\bibnamefont {Bersin}}, \bibinfo {author} {\bibfnamefont
  {C.}~\bibnamefont {M{\"u}ller}}, \bibinfo {author} {\bibfnamefont
  {L.}~\bibnamefont {McGuinness}}, \bibinfo {author} {\bibfnamefont
  {F.}~\bibnamefont {Jelezko}}, \bibinfo {author} {\bibfnamefont {R.~L.}\
  \bibnamefont {Walsworth}}, \bibinfo {author} {\bibfnamefont {H.}~\bibnamefont
  {Park}}, \ and\ \bibinfo {author} {\bibfnamefont {M.~D.}\ \bibnamefont
  {Lukin}},\ }\href {\doibase 10.1126/science.aad8022} {\bibfield  {journal}
  {\bibinfo  {journal} {Science}\ }\textbf {\bibinfo {volume} {351}},\ \bibinfo
  {pages} {836} (\bibinfo {year} {2016})}\BibitemShut {NoStop}%
\bibitem [{\citenamefont {Jiang}\ \emph {et~al.}(2009)\citenamefont {Jiang},
  \citenamefont {Hodges}, \citenamefont {Maze}, \citenamefont {Maurer},
  \citenamefont {Taylor}, \citenamefont {Cory}, \citenamefont {Hemmer},
  \citenamefont {Walsworth}, \citenamefont {Yacoby}, \citenamefont {Zibrov},\
  and\ \citenamefont {Lukin}}]{Jiang2009}%
  \BibitemOpen
  \bibfield  {author} {\bibinfo {author} {\bibfnamefont {L.}~\bibnamefont
  {Jiang}}, \bibinfo {author} {\bibfnamefont {J.~S.}\ \bibnamefont {Hodges}},
  \bibinfo {author} {\bibfnamefont {J.~R.}\ \bibnamefont {Maze}}, \bibinfo
  {author} {\bibfnamefont {P.}~\bibnamefont {Maurer}}, \bibinfo {author}
  {\bibfnamefont {J.~M.}\ \bibnamefont {Taylor}}, \bibinfo {author}
  {\bibfnamefont {D.~G.}\ \bibnamefont {Cory}}, \bibinfo {author}
  {\bibfnamefont {P.~R.}\ \bibnamefont {Hemmer}}, \bibinfo {author}
  {\bibfnamefont {R.~L.}\ \bibnamefont {Walsworth}}, \bibinfo {author}
  {\bibfnamefont {A.}~\bibnamefont {Yacoby}}, \bibinfo {author} {\bibfnamefont
  {A.~S.}\ \bibnamefont {Zibrov}}, \ and\ \bibinfo {author} {\bibfnamefont
  {M.~D.}\ \bibnamefont {Lukin}},\ }\href {\doibase 10.1126/science.1176496}
  {\bibfield  {journal} {\bibinfo  {journal} {Science}\ }\textbf {\bibinfo
  {volume} {326}},\ \bibinfo {pages} {267} (\bibinfo {year}
  {2009})}\BibitemShut {NoStop}%
\bibitem [{\citenamefont {Robledo}\ \emph {et~al.}(2011)\citenamefont
  {Robledo}, \citenamefont {Childress}, \citenamefont {Bernien}, \citenamefont
  {Hensen}, \citenamefont {Alkemade},\ and\ \citenamefont
  {Hanson}}]{Robledo2011}%
  \BibitemOpen
  \bibfield  {author} {\bibinfo {author} {\bibfnamefont {L.}~\bibnamefont
  {Robledo}}, \bibinfo {author} {\bibfnamefont {L.}~\bibnamefont {Childress}},
  \bibinfo {author} {\bibfnamefont {H.}~\bibnamefont {Bernien}}, \bibinfo
  {author} {\bibfnamefont {B.}~\bibnamefont {Hensen}}, \bibinfo {author}
  {\bibfnamefont {P.~F.~A.}\ \bibnamefont {Alkemade}}, \ and\ \bibinfo {author}
  {\bibfnamefont {R.}~\bibnamefont {Hanson}},\ }\href
  {http://dx.doi.org/10.1038/nature10401} {\bibfield  {journal} {\bibinfo
  {journal} {Nature}\ }\textbf {\bibinfo {volume} {477}},\ \bibinfo {pages}
  {574} (\bibinfo {year} {2011})}\BibitemShut {NoStop}%
\end{thebibliography}%
\end{document}


\title{Supplemental Material for \\ ``Near-Infrared-Assisted Charge Control and Spin Readout\\ of the Nitrogen-Vacancy Center in Diamond''}

\author{David A.\ Hopper}
	\affiliation{Quantum Engineering Laboratory, Department of Electrical and Systems Engineering, University of 			Pennsylvania, Philadelphia, PA 19104 USA}
    \affiliation{Department of Physics, University of Pennsylvania, Philadelphia, PA 19104 USA}
\author{Richard R. Grote}
	\affiliation{Quantum Engineering Laboratory, Department of Electrical and Systems Engineering, University of 			Pennsylvania, Philadelphia, PA 19104 USA}
\author{Annemarie L.\ Exarhos}    
	\affiliation{Quantum Engineering Laboratory, Department of Electrical and Systems Engineering, University of 			Pennsylvania, Philadelphia, PA 19104 USA}   
\author{Lee C.\ Bassett}
	\email[Corresponding author.\\ Email address: ]{lbassett@seas.upenn.edu}
	\affiliation{Quantum Engineering Laboratory, Department of Electrical and Systems Engineering, University of 			Pennsylvania, Philadelphia, PA 19104 USA}      
\date{\today}

\date{\today}

\maketitle


\section{Experimental Set Up}
The excitation sources used are as follows: a \SI{532}{\nano\meter} continuous-wave diode-pumped solid state laser (gem 532, Laser Quantum), a \SI{592}{\nano\meter} continuous wave external cavity fiber laser (VFL-592, MPB communications, Inc.), and a single-shot $\sim$~\SI{10}{\pico\second}-pulse supercontinuum source (WhiteLase SC-400, Fianium with Ultra Pod option) band-filtered to 900-\SI{1000}{\nano\meter}.  Acousto-optic modulators (AOMs, 1250C, Isomet) are used to temporally gate the visible beams. The \SI{532}{\nano\meter} AOM is set up in a double-pass configuration which improves the extinction to $>$\SI{60}{\decibel} to prevent cycling of the charge state during high fidelity readouts. The supercontinuum source outputs NIR pulses at a repetition rate of \SI{40}{\mega\hertz} that can be fully gated on demand using a single-shot output option. Excitation beams are collimated and co-aligned on a fast steering mirror (FSM, OIM101, Optics in Motion) and imaged onto the back of an objective (Olympus 100x NA 0.9) using a 4$f$ lens configuration. Photoluminescence is collected through the same objective and focused onto a \SI{50}{\micro\meter} diameter core multi-mode fiber that is connected to a low dark count (\SI{20}{\counts\per\second}) single-photon avalanche diode (SPAD) (Count-20C-FC, Laser Components). A data acquisition card (DAQ-6323, National Instruments) serves as the master clock, controlling the millisecond timing and counter input for the SPAD. An arbitrary waveform generator (AWG520, Tektronix) with a sample rate of \SI{1}{\giga\samples\per\second} controls the timing on the optical dynamics time scales and is triggered by the data acquisition card.

\begin{figure}
\includegraphics[width=\linewidth]{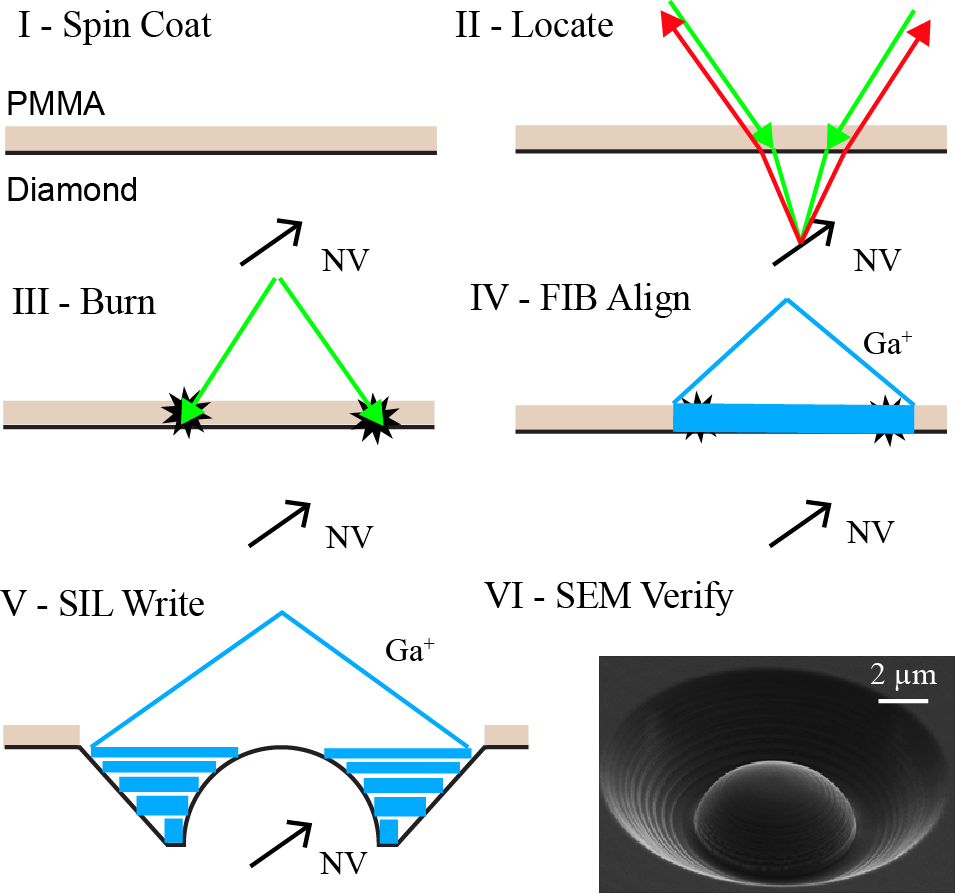}
\caption{SIL Fabrication Process (I) Spin coat diamond with 500 nm of PMMA A8 to act as an \textit{in situ} alignment medium. (II) Image NV centes to find suitable candidates ($\sim$\SI{3}{\micro\meter} below surface) (III) Burn \SI{8}{\micro\meter} square marks centered on the NV ( $\sim$\SI{100}{\milli\watt} of \SI{532}{\nano\meter}) for 7 seconds to visibly indicate the NV location. (IV) Locate the markers in the FIB SEM, raster off the PMMA in the vicinity to be etched. (V) Etch the SIL using concentric circles of increasing depth. (VI) Verify the etch in the SEM and subsequently in the confocal microscope. \label{fig:SILfabrication}}
\end{figure}

\section{Device Fabrication}
All NVs measured are contained within solid immersion lenses (SILs) fabricated on the diamond surface with a focused ion beam (FIB). The alignment procedure is shown schematically in Fig.~\ref{fig:SILfabrication}. We first spin coat a \SI{500}{\nano\meter} layer of PMMA A8 resist (MicroChem Nano PMMA) on the diamond surface, and do not bake the resist. The sample is then mounted in the experimental setup described above using a rotation compensating mount (Thorlabs KM100T), which allows for precise tilt correction of the sample relative to the optical axis. We ensure the diamond surface is perpendicular to the optical axis by moving the FSM over \SI{80}{\micro\meter} and checking the relative height difference by observing the focal position of the green laser in a camera. The rotation compensating mount is used to offset any tilt in the $x$ or $y$ axis. We achieve a tilt angle that is with in $\pm$~\SI{0.1}{\degree} of the optical axis using this technique. After tilt correction, we identify NVs of a desired depth below the diamond surface, typically \SI{3}{\micro\meter} for this work, by imaging through the PMMA layer, which is transparent in the visible spectrum. We determine the NV depth by finding the relative distance of the NV from the surface of the diamond using PL as an indicator. Following the identification of a candidate NV, we focus and align to our NV, move the beam focus to the surface, increase the laser power to $\sim$\SI{100}{\milli\watt} of \SI{532}{\nano\meter}, and then burn five holes into the PMMA, 4 corners of an \SI{8}{\micro\meter} box surrounding the center of the NV, and one directly above the NV, by dwelling at each point for 7 seconds to ensure a mark is made.  This time can be adjusted higher or lower based on the available green power. However, longer times lead to drift during the burn process which will reduce the yield.

After a suitable number of candidate NVs are marked and aligned, we sputter a discharge layer of AuPd on top of the PMMA. The sample is then loaded in a dual beam scanning electron microscope and focused ion beam (SEM/FIB, Strata D235, FEI), taking care to securely ground the AuPd discharge layer to prevent drift during fabrication due to charging. The PMMA burn marks are located in the SEM, and a SIL is fabricated directly through the PMMA layer in a method similar to Ref. \cite{Jamali2014}. Following the fabrication, the PMMA and AuPd layer is removed by sonication in Microposit Remover 1165 (MicroChem).  The gallium-implanted diamond layer left by FIB milling is removed by etching \SI{80}{\nano\meter} of diamond using an Ar/Cl ICP/RIE etch (Trion Phantom ICP, Pressure: 10 mT, ICP: \SI{400}{\watt}, RIE: \SI{300}{\watt}, Ar: 12 sccm, Cl$_2$: 20 sccm) \cite{Lee2008}. Finally, the sample is cleaned in Nano-strip (Cyantek) at \SI{70}{\celsius} for \SI{20}{\minute}, followed by a soft-O$_2$ plasma clean for 15 minutes (Anatech SCE-108 Barrel Asher, RF power = \SI{30}{\watt}) to remove any graphite layer and oxygen terminate the surface. We then verify the alignment and fabrication by imaging in the confocal set up.  For the NVs reported in this work, we find that the collection efficiency is improved by a factor of 6 and the excitation efficiency by a factor of 10 from saturation curve measurements.

\section{Charge State Measurements}
\begin{figure}[t]
\includegraphics[scale=1]{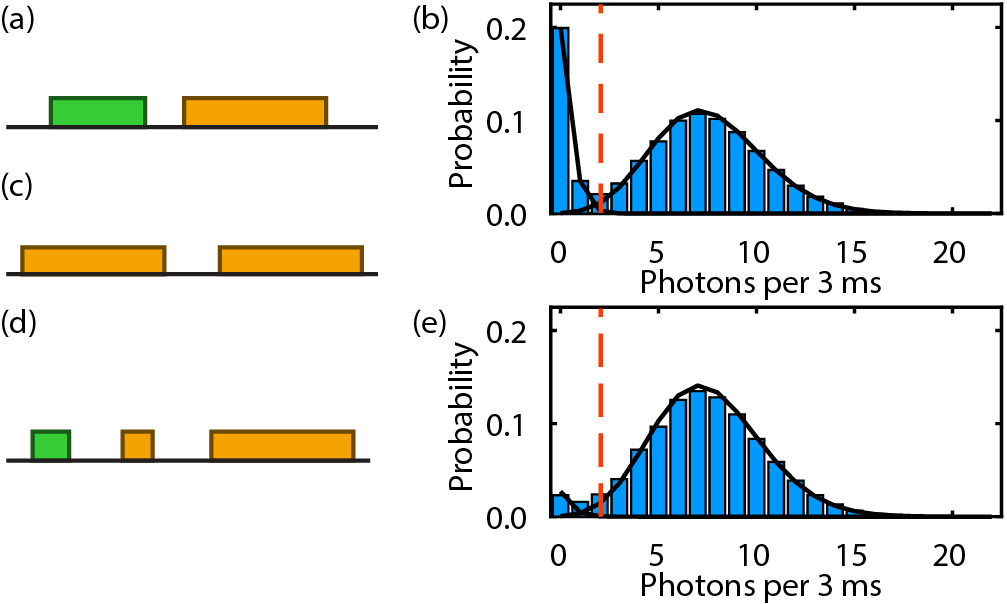}
\caption{Charge State Measurements (a) Pulse sequence used to measure the charge state with high-fidelity and the monochromatic initialization into NV$^-$. (b) Measured photon-counting histogram (bars) from (a). Curves are fits to two Poissonian functions that determine the mean counts and relative populations of the dark and bright states. (c) Pulse sequence for measuring the non-destructivity of the charge state measurement. (d) Pulse sequence for verifying post selection into NV$^-$. (e) Results of post-selected charge state measurement conditioned on one or more photons detected during the verification step. }
\label{fig:CSExample}
\end{figure}

Charge state measurements are conducted using \SI{592}{\nano\meter} light with a power setting between 75-\SI{300}{\nano\watt}, based on the desired trade-off between measurement speed and destructivity. The exact power depends heavily on sample collection and excitation efficiency, but can easily be found by performing a saturation curve and operating $\sim$3 orders of magnitude below saturation. Ionization and recombination under visible illumination limits the speed at which the charge can be readout due to the added destructivity. The power ($P_{592}$) and duration counting photons ($\tau_{R}$) are chosen to maximize the signal-to-noise ratio (SNR) of detected photons ($\gamma_{B}/\gamma_{D}\gg1$) without destroying the state ($\gamma_{Ion}\tau_R, \gamma_{Rec}\tau_R \ll 1$). A consideration of these constraints allows for further categorization of the measurement into three regimes that are useful for different applications: high fidelity and measurement speed, high fidelity and non-destructivity, and high fidelity verification of NV$^-$ by post-selection. 

An example of a typical high-fidelity/high-speed measurement is depicted in Fig.~\ref{fig:CSExample}(a), where a relatively high power($P_{592}$=\SI{220}{\nano\watt}) and shorter readout duration($\tau_R$=\SI{3}{\milli\second}) leads to a single-shot charge state fidelity of $\mathcal{F}_c = 99.1\pm0.4\%$ with a threshold of 3 photons (Fig.~\ref{fig:CSExample}(b)). The fidelity is defined according to $\mathcal{F}_c=1-(\varepsilon_0+\varepsilon_-)/2$, where $\varepsilon_i$ is the error probability in measuring charge-state $i$.  For example, $\varepsilon_0 =P(-|0)$ is the probability of identifying the charge state NV$^-$ (detecting a photon number above the threshold) given that the state was actually NV$^0$.

The non-destructivity of the measurement can be increased at the expense of measurement speed by reducing the power and increasing the measurement time.  This allows for the observation of individual charge state transitions and the direct measurement of rates as in Fig.~3(c-d) of the main text. The figure of merit then includes both the charge state fidelity and the non-destructivity, $\mathcal{F}_D$, which is the probability that the charge state is unaltered by the measurement. By repetitively measuring the charge state as in Fig.~\ref{fig:CSExample}(c), we can extract both $\mathcal{F}_C$ and $\mathcal{F}_D$.  For $P_{592}$ = \SI{75}{\nano\watt} and $\tau_R$ = \SI{15}{\milli\second}, we measure $\mathcal{F}_c = 99.3\pm0.4\%$ with a non-destructivity of 96.1$\pm0.2\%$ and 99.0$\pm0.1\%$ for NV$^-$ and NV$^0$, respectively. The charge dependence of $\mathcal{F}_D$ results from the fact that ionization is $\sim$5 times more likely than recombination in this scenario. 

Finally, by reducing $\tau_R$ such that we obtain $\approx1$ photon during a readout window, we can efficiently post select into NV$^-$ based on the presence of one or more photons during the verification readout (Fig.~\ref{fig:CSExample}(d)). An example of this  post selection measurement followed by a full high fidelity charge state measurement is depicted in Fig.~\ref{fig:CSExample}(e).  Using a \SI{420}{\micro\second} verification step and a $\tau_R$=\SI{3}{\milli\second} high fidelity measurement, and a common $P_{592}$=\SI{220}{\nano\watt}, we can post-select into NV$^-$ with fidelity $\mathcal{F}_{PS}$ = 96.8$\pm0.4\%$.  This value agrees with the expected errors due to mis-identification and ionization during verification as discussed above.

\section{Phenomenological Model for Steady-State Charge}

\begin{figure}[]
\includegraphics[width=\linewidth]{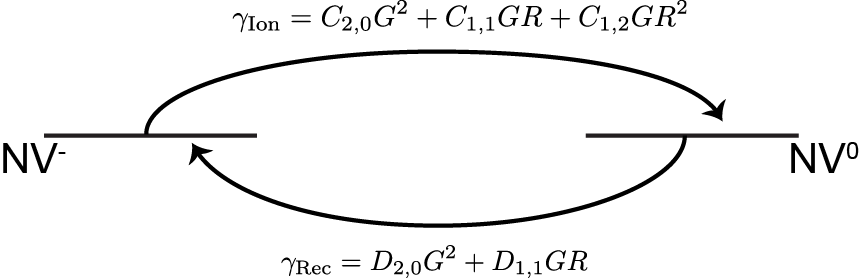}
\caption{Schematic of the two-level rate-equation model used to fit the steady-state charge distributions in the main text. Rate coefficients correspond to the transitions presented in Fig.~3(e) of of the main text.}
\label{fig:PhenomModel}
\end{figure}

To model the steady-state charge distributions measured as a function of visible and NIR power in Figs.~2(b) and 3(a) of the main text, we compress the six-level system depicted in Fig.~3(e) of the main text into a two-level phenomenological model that accounts for the important nonlinear absorption terms that drive ionization/recombination in the presence of visible$+$NIR illumination.  The terms we include involve at least one visible photon, as shown in Fig.~\ref{fig:PhenomModel}, where $G$ and $R$ refer to the \SI{532}{\nano\meter} and NIR power, respectively, and the coefficients $\{C_{m,n}, D_{m,n}\}$ are defined in the main text.  We do not include the multiphoton NIR-only transitions indicated in Fig.~3(e) since their rates (Fig.~3(d)) are much slower than for the corresponding visible$+$NIR processes at the same NIR power.  This agrees with the many-orders-of-magnitude smaller cross-sections of virtual two photon absorption of both charge states compared to single-photon processes \cite{Ivanov2013}.

With these approximations, the master equation governing the charge-state evolution becomes
\begin{widetext}
\begin{equation}
\frac{d}{dt}\begin{pmatrix}
p_{-} \\
p_0
\end{pmatrix}
=
\begin{pmatrix}
-\gamma_{\text{Ion}} & \gamma_{\text{Rec}} \\
\gamma_{\text{Ion}} & -\gamma_{\text{Rec}}
\end{pmatrix}
\begin{pmatrix}
p_{-} \\
p_0
\end{pmatrix}
=
\begin{pmatrix}
- C_{2,0}G^2 - C_{1, 1}GR - C_{1, 2}GR^2 & D_{2, 0} G^2 + D_{1, 1}GR \\
C_{2,0}G^2 + C_{1, 1}GR + C_{1, 2}GR^2 & -D_{2, 0} G^2 - D_{1, 1}GR
\end{pmatrix}
\begin{pmatrix}
p_{-} \\
p_0
\end{pmatrix},
\end{equation}
\end{widetext}
We solve for the steady-state solution, along with the condition the population must be conserved, and obtain expressions for $p_-$ and $p_0$. Below we show the analysis for $p_-$, since it is directly related to the fits in the main text. The solution can be written in the form
\begin{equation}
p_- = \gamma \frac{1+\alpha R}{1+\delta R + \beta R^2},
\end{equation}
where
\begin{subequations}
\begin{align}
\alpha & = \frac{D_{1, 1}}{D_{2, 0}G}, \\
\beta & = \frac{C_{1, 2}}{(C_{2, 0} + D_{2, 0})G},\\
\gamma & = \frac{D_{2, 0}}{D_{2,0} + C_{2, 0}}, \\
\delta & = \frac{C_{1, 1} + D_{1, 1}}{(C_{2, 0} + D_{2, 0})G}.
\end{align}
\end{subequations}
Note that $\alpha$, $\beta$, and $\delta$ scale with $1/G$.  This is expected since they depend implicitly on the population of internal metastable states. In a full solution of the six-level model (not shown), the parameters  $\{C_{m,n}, D_{m,n}\}$ also depend non-trivially on $R$ and $G$ since they account for both the (constant) absorption cross sections of various processes and the (power-dependent) occupation probabilities of the levels in the model.  Nonetheless, it is a reasonable approximation to treat them as constant parameters across the range of powers considered here, particularly as discussed above when multiphoton NIR-only transitions are not playing a major role.

To fit the data in Figs.~2(b) and 3(a) of the main text, the 4 parameters $(\alpha,\beta,\gamma,\delta)$ are allowed to vary independently. Note that even \SI{592}{\nano\meter} light does not excite NV$^0$ to lowest order, at room temperature there is non-zero absorption due to an anti-Stokes shift \cite{Aslam2013}, so the recombination coefficients $D_{2,0}$ and $D_{1,1}$ are nonzero even for the fits to the \SI{592}{\nano\meter}$+$NIR data in Fig.~3(a).

\begin{figure}[t]
\includegraphics[width=\linewidth]{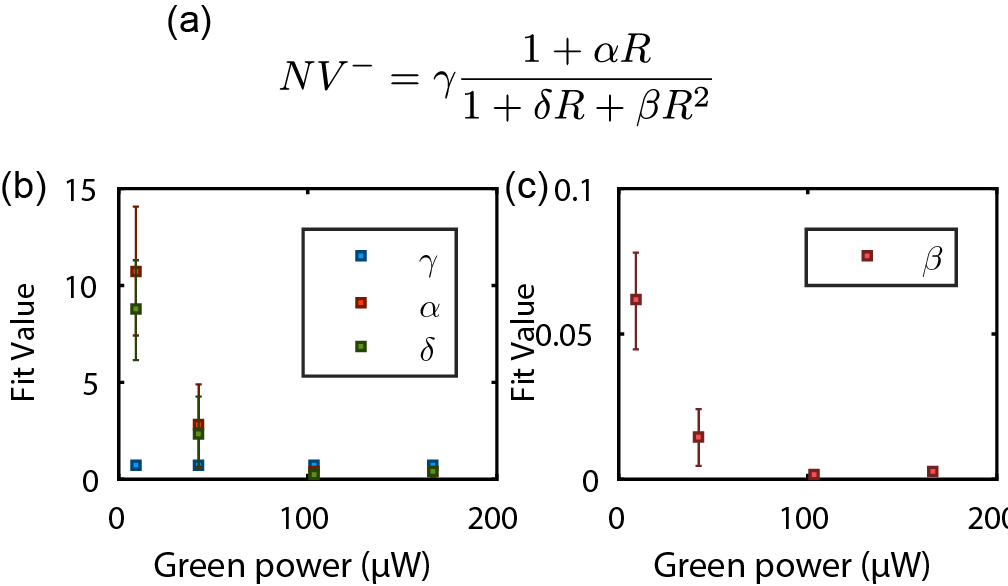}
\caption{Results of the Phenomenological model fit (a) The actual fit equation we use, with simplified coefficients (b) the first three coefficients (c) The last coefficient, which is about two orders of magnitude smaller (signifying the weakness of the singlet ionization mechanism). Error bars are 65$\%$ confidence. \label{fig:PhenomFit}}
\end{figure}

In Fig.~2 of the main text, we swept $R$ for over four different settings of $G$, varying across above the PL saturation value of $P_\mathrm{sat}=42$~\si{\micro\watt}. We independently fit the results of four green power slices using this model (only three are shown in the main text for clarity but all are presented here).  The best-fit parameters are plotted in Fig.~\ref{fig:PhenomFit}.  We observe the expected scaling with $1/G$ in the best-fit parameters $\alpha$, $\beta$, and $\delta$, justifying our approximation of fixing the remaining rate coefficients in the model.

Through this parameterization, we have reduced the number of free parameters in the model from five ($C_{1,1}, D_{1,1}, C_{2,0}, D_{2,0}$, and $C_{1,2}$) to four that uniquely control the model's dependence on $R$.  From the fit results, we can back out the relative strengths of many of the underlying rate coefficients.  For example, the best-fit value of $\gamma$ corresponds to the steady-state value of $p_-$ under visible illumination only.  For \SI{532}{\nano\meter} illumination, the observed value of $p_-= 78$\% therefore implies that $R_{1,1}/C_{1,1}=3.5$.

Similarly, the ratio $\alpha/\delta$ can be writen in the form
\begin{align}
\frac{\alpha}{\delta} = \frac{D_{1, 1}}{(D_{1, 1} + C_{1, 1})}\frac{(D_{2, 0} + C_{2, 0})}{D_{2, 0}} \nonumber \\
= \frac{D_{1, 1}}{(D_{1, 1} + C_{1, 1})}\frac{1}{\gamma}.
\end{align}
We can therefore calculate the relative strength of the NIR-assisted ionization/recombination transitions from the best-fit values of $\alpha$, $\gamma$, and $\delta$.  We find $D_{1, 1}/C_{1,1} = 6.69\pm0.04$. The fact that this ratio is larger than for the analogous process for two \SI{532}{\nano\meter} photons intuitively explains the initial enhancement of $p_-$ as a function of $R$, as recombination becomes even more likely than ionization.

Finally, We can extract the relative strength of two-NIR-photon singlet ionization is compared to the competing process for excited state recombination in a similar manner:
\begin{equation}
\frac{\beta}{\alpha} = \frac{C_{1,2}}{C_{2, 0} + D_{2, 0}}\frac{D_{2,0}}{D_{1,1}}
= \frac{C_{1, 2}}{D_{1,1}}\gamma,
\end{equation} 
from which we find $C_{1, 2}/D_{1,1} = 7.4\pm0.3\times10^{-3}$. This much smaller strength is not surprising considering the small optical cross section of the singlet.  Nonetheless, singlet ionization plays a dominant role in the charge-state dynamics at high NIR powers due to the quadratic scaling with $R^2$.

\section{Spin Polarization}
We have not directly measured the spin polarization of the enhanced NV$^-$ initialization protocol. However, we expect that the spin polarization should be similar to that obtained by visible illumination alone (typically, $\sim$80-90\%). The observed charge switching rates are on the order of \SI{10}{\kilo\hertz}, whereas the spin polarization is typically limited by the singlet lifetime, which is \SI{4}{\mega\hertz}. Even when using lower-power green illumination, where the spin polarization may be slightly slower, a safe lower bound is \SI{500}{\kilo\hertz}, equating to a \SI{2}{\micro\second} polarization time. This implies that the spin should be polarized two orders of magnitude faster than the charge state is changing, which should provide ample time for the spin to remain polarized with the enhanced NV$^-$ population.

\begin{figure}
\includegraphics[width=\linewidth]{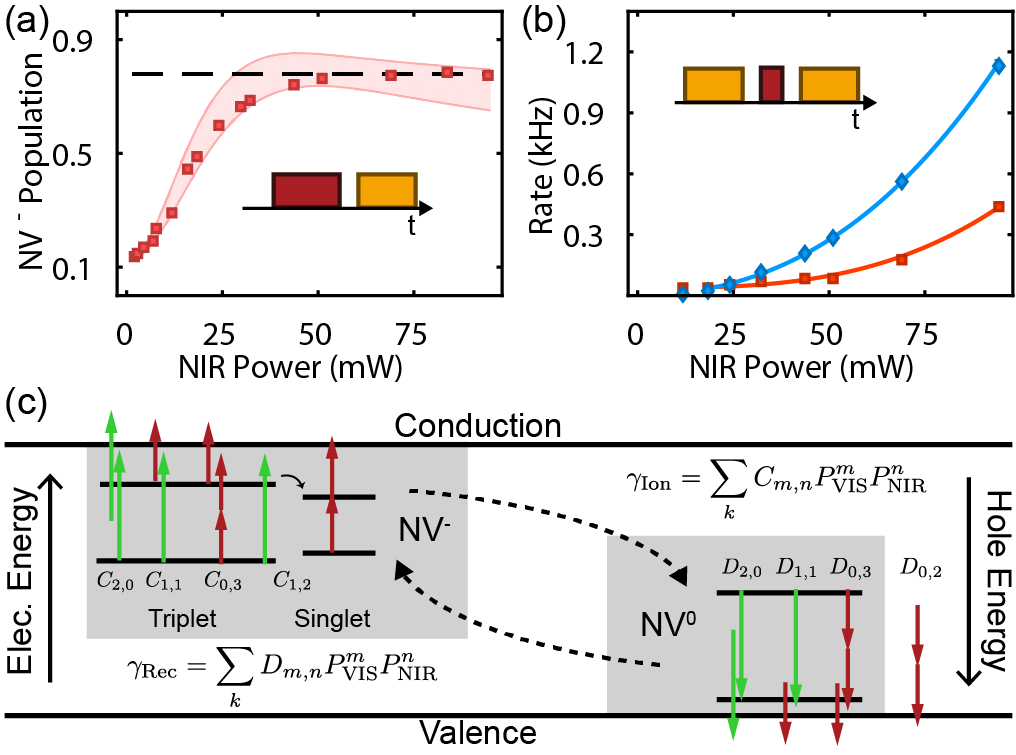}
\caption{Compiliation of the NIR only data (a) Recreated NIR NV$^-$ population for a fixed \SI{10}{\milli\second} illumination period from the main text. (b) Measured ionization (red squares) and recombination rates (blue diamonds) due to NIR illumination alone. (c) The full allowed charge transitions including the NIR only transitions. $D_{0, 2}$ corresponds to a two photon recombination process that is required to fit the data, see Fig~\ref{fig:NIRrecombFit} below.}
\label{fig:NIROnly}
\end{figure}

\section{NIR-only analysis}
We performed a rate measurement for the NIR only illumination in a similar manner to the green+NIR measurement of the main text. The resulting NIR population after a \SI{10}{\milli\second} illumination duration and the corresponding rates are depicted in Fig.~\ref{fig:NIROnly}(a-b). The magnitude of the NIR-only rates are several orders of magnitude smaller than for the case of coupled visible$+$NIR illumination across the full range of powers we consider in this work, which justifies excluding them from the phenomenological master-equation model discussed above. This difference in magnitude is not surprising considering the fact that these NIR-only transitions rely on virtual multi-photon transitions with much smaller cross sections than for the sequential single-photon absorption responsible for the coupled rates.   Fig.~\ref{fig:NIROnly}(c) includes a complete diagram of all the multiphoton charge transitions we have detected, including the NIR-only processes. 

\begin{figure}
\includegraphics[width=\linewidth]{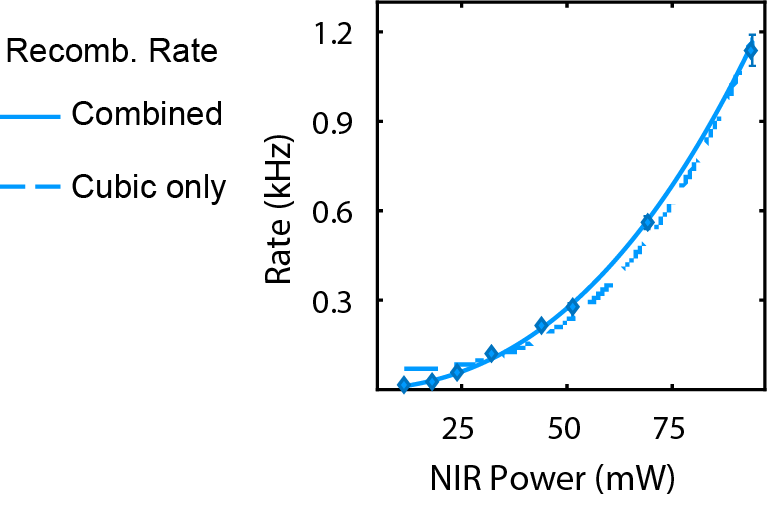}
\caption{Comparison of a cubic only and cubic + quadratic fit of the recombination rate for NIR illumination alone. Error bars are smaller than the symbols except where plotted. \label{fig:NIRrecombFit}}
\end{figure}

Although the NIR-only processes are too weak to influence the multicolor dynamics that are the focus of the main text, we note an interesting and unexpected outcome in the power dependence of the NIR-only recombination rate. According to the level structure in Fig.~\ref{fig:NIROnly}(c), the lowest-order ionization/recombination process available in the presence of NIR light alone should involve three photons and hence the ionization/recombination rates should depend on $R^3$.  We find, however, that an additional quadratic term is needed to yield a satisfactory fit of the recombination rate measurements.  Fig.~\ref{fig:NIRrecombFit} shows the data together with fits consisting of a cubic term only compared to that used in the text which includes a quadratic term. The actual polynomial function used to fit the data in Figs.~3(c-d) of the main text is
\begin{align}
\gamma_{R/I} = aR^3 + bR^2 + c,
\end{align}
where the offset $c$ is required to account for small but nonzero destructivity of the charge-state measurement. For the ionization fit, we force b to be zero, and the fit parameters are well constrained, whereas if we do the same for the recombination rate, the fit is inconsistent with the data. This observation deserves further study in the future. We tentatively propose that the quadratic-in-$R$ recombination process might result from two-photon ionization of nearby substitutional nitrogen impurities and subsequent electron capture by the NV.  If this hypothesis is true, we should expect to observe variations between different NV centers based on their local environments. Best-fit parameters for both fits are listed in Table~\ref{tab:Fit}. The $R^2$ fit-statistic values for the fits are 0.991 and 0.992 for recombination and ionization respectively.

\setlength{\extrarowheight}{5pt}
\begin{table}
\caption{Results of the cubic fit for Ionization, and the cubic and quadratic fit for recombination.}
\begin{tabular}{|l|l|l|l|}
\hline
Process & a (\si{\kilo\hertz\per\milli\watt\cubed}) & b (\si{\kilo\hertz\per\milli\watt\squared}) & c (\si{\kilo\hertz}) \\
\hline
Ion & $4.7\pm0.4\times10^{-7}$ & 0 (fixed) & $0.039\pm0.014$ \\
\hline
Rec &$5.1\pm3.7\times10^{-7}$ & $8.4\pm2.1\times10^{-5}$ & $1\pm0.1\times10^{-7}$ \\
\hline
\end{tabular}
\label{tab:Fit}
\end{table}

In contrast to the case of visible$+$NIR excitation in Fig.~3(a), the NIR-only measurements do not generally reflect the steady-state value of $p_-$, due to the slow underlying rates.  Instead, what we measure is a non-equilibrium distribution that results from competition between recombination/ionization due to NIR alone (which are too slow to yield a steady-state population at low NIR powers) and the backaction of the charge state measurement (which is optimized for readout fidelity rather than non-destructivity).  Still, given the extracted ionization and recombination rates as a function of $R$ from Fig.~\ref{fig:NIROnly}(b), we can predict the charge distributions that are expected from the experiment reported in Fig.~3(a) of the main text.

The evolution of the population vector $\mathbf{p} = \bigl( \begin{smallmatrix} p_- \\ p_0
\end{smallmatrix} \bigr)$ during a single cycle of the population measurement is
\begin{equation}
\mathbf{p'} = [M(R)\cdot D]\mathbf{p},
\label{eq:pEvolutionNIR}
\end{equation}
where $M(R)$ is the evolution matrix due to NIR light alone and $D$ captures the effect of the destructive readout step.  The evolution matrix $M(R)$ can be found by integrating the master equation for the interaction time $t_R=10$~\si{\milli\second} using the ionization/recombination rates we directly measured in Fig.~\ref{fig:NIROnly}(b) for each corresponding NIR power setting.  By characterizing the readout destructivity as described earlier, we extract
\begin{equation}
D=
\begin{pmatrix}
0.65 & 0.05 \\
0.35 & 0.95
\end{pmatrix}.
\end{equation}
The measured populations, therefore, correspond to the equilibrium state of the evolution described by eqn.~(\ref{eq:pEvolutionNIR}), which is simply the normalized eigenvector of $[M(R)\cdot D]$ corresponding to its unity eigenvalue.

This analysis is used to produce the simulation plotted as a shaded region in Fig.~3(b) of the main text. The shaded region corresponds to the confidence interval corresponding to the uncertainty in the underlying polynomial fits to the ionization/recombination rates (Table~\ref{tab:Fit}). Minor disagreements between the simulation and measurements might result from shifts in the microscope's optical alignment that alter the NIR intensity for a given power setting, since the data in Figs.~3(a) and \ref{fig:NIROnly}(b) were taken at different times.

Notably, our observation of NV$^-$ enhancement at higher NIR power contradicts the conclusions from earlier two-photon absorption measurements in nanodiamonds \cite{Ivanov2013}.  We believe that differences in the host material, NIR wavelength, power, and pulse duration might explain the discrepancy, together with the complicated dynamics at high NIR powers due to multiple competing nonlinear processes, but this topic also warrants further study in the future.

\section{Population Transfer Matrix Model for Spin-to-Charge Conversion}

To quantify the performance of the multi-SCC process, we use a six level model taking into account the $m_s=0, -1$ spin sublevels of the NV$^-$ triplet, the metastable single ground state, and a single NV$^0$ state. We ignore the singlet excited state since its $\sim$\SI{1}{\nano\second} lifetime is both much shorter than the metastable singlet ground state and much longer than the $\sim$\SI{10}{\pico\second} duration of a NIR pulse.  Similarly, we ignore the NV$^0$ excited state since its dynamics are implicit in the values of the recombination coefficients.

\begin{figure}
\includegraphics[width=\linewidth]{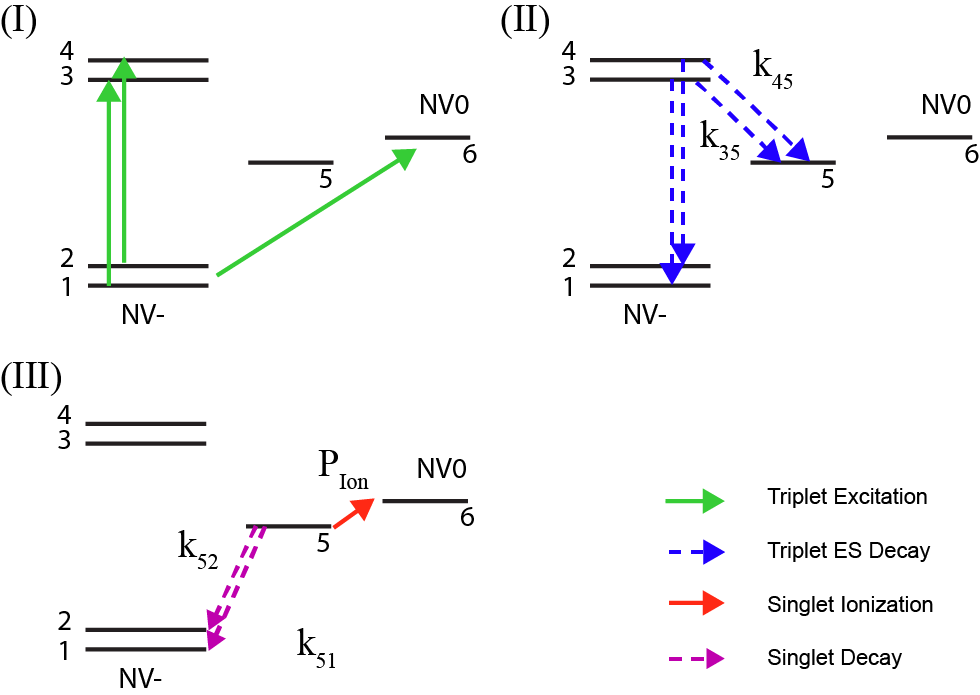}
\caption{Outline of the 6-level population transfer matrix model for multi-SCC (I) excite the triplet ground state. (II) Allow the excited state to decay. (III) Attempt to ionize the singlet and decay to the ground state. The probability is conserved leaving any state.}
\label{fig:TMM}
\end{figure}

A depiction of these levels is presented in Fig.~\ref{fig:TMM}. A single SCC step is broken into three distinct regions. The first corresponds to the excitation of the triplet ((I) in Fig.~\ref{fig:TMM}), which also allows for ionization. While the physical ionization process is sequential two-photon absorption through the triplet excited states (3 and 4), we are only interested in the resulting distribution of populations between the triplet and NV$^0$ manifolds, so we combine the sequential absorption into a single step. The system is then allowed to relax either to the ground state or the singlet manifold through the ISC, which we define by the branching ratios k$_{35}$ and k$_{45}$ for the probability of $m_s=0$ and $m_s=\pm1$ respectively to decay into the singlet (Fig.~\ref{fig:TMM}(II)). As in step (I), we ignore the intermediate dynamics of the singlet excited state since it relaxes on a much faster timescale ($\sim$\SI{1}{\nano\second}) than the metastable singlet ground-state lifetime ($\sim$\SI{200}{\nano\second}).

In the ideal case, the system would only be excited once, however we cannot guarantee this due to long AOM turn-on times. Therefore the effective ISC transition rates we measure likely correspond to a few optical cycles. After the triplet excited state has fully decayed, we attempt to ionize the singlet with a train of NIR pulses, which we treat as a single step (Fig.~\ref{fig:TMM}(III)).  Any population that is not ionized decays into the ground state, with the branching ratios k$_{51}$ and k$_{52}$. 

To construct the master equation, we define a population vector
\begin{align}
p = \begin{pmatrix}
p1 \\ p2 \\ p3 \\ p4 \\ p5 \\ p6
\end{pmatrix},
\end{align}
and matrix representations for the steps I--III:
\begin{equation}
M_\mathrm{I} = 
\begin{pmatrix}
0 & 0 & 0 & 0 & 0 & 0 \\
0 & 0 & 0 & 0 & 0 & 0 \\
P_{exc} & 0 & 0 & 0 & 0 & 0 \\
0 & P_{exc} & 0 & 0 & 0 & 0 \\
0 & 0 & 0 & 0 & 0 & 0 \\
P_{ion} & P_{ion} & 0 & 0 & 0 & 1 \\
\end{pmatrix},
\end{equation}
\begin{equation}
M_\mathrm{II} = 
\begin{pmatrix}
1 & 0 & 1-k_{35} & 0 & 0 & 0 \\
0 & 1 & 0 & 1-k_{45} & 0 & 0 \\
0 & 0 & 0 & 0 & 0 & 0 \\
0 & 0 & 0 & 0 & 0 & 0 \\
0 & 0 &  k_{35} & k_{45} & 0 & 0 \\
0 & 0 & 0 & 0 & 0 &1\\
\end{pmatrix},
\end{equation}
\begin{equation}
M_\mathrm{III} = 
\begin{pmatrix}
1 & 0 & 0 & 0 & k_{51} & 0 \\
0 & 1 & 0 & 0 & k_{52} & 0 \\
0 & 0 & 0 & 0 & 0 & 0 \\
0 & 0 & 0 & 0 & 0 & 0 \\
0 & 0 & 0 & 0 & 0 & 0 \\
0 & 0 & 0 & 0 & P_{\text{sing}} & 0 \\
\end{pmatrix}.
\end{equation}
Note that the columns must sum to 1 to conserve probability, and that only the states 1, 2, and 6 are stable (and thus never decay). The expected populations after an arbitrary number of repeats, $N$, are therefore given by
\begin{align}
p_\mathrm{final} = \left(M_\mathrm{III}\times M_\mathrm{II}\times M_\mathrm{I}\right)^Np_0,
\end{align}
and the first two elements of $p_\mathrm{final}$ constitute the resulting NV$^-$ population.

We perform a joint fit of this model to both spin-initialization datasets from Fig.~4(c) in the main text, accounting also for incomplete spin and charge initialization in the definition of $p_0$. The charge-initialization is known directly from calibration measurements, while the spin-initialization remains a free parameter.  Separate measurements of the green shelving pulse alone fix the ionization probability during excitation to $P_\mathrm{ion}=0.5$\%. This leaves six free parameters in the joint fit, whose best-fit values and uncertainties are listed in Table~\ref{tab:SCCFit}.

\section{Spin-Readout Figures of Merit}
Below we derive the SNR of spin readout due to the combined effects of imperfect SCC efficiency and shot noise in the charge-state readout.  We also consider how to optimize the readout parameters to yield the best performance of time-averaged spin measurements. In general, the noise in the final signal depends on the initial spin state.  Spin-projection noise plays a role for superposition states, but there can also be a large difference in the observed signal variance even for spin eigenstates due to asymmetries in the SCC process and Poissonian statistics of the charge-state readout. For that reason we define our figure-of-merit SNR corresponding to a \emph{differential measurement} of the signals resulting from spins prepared in the eigenstates $m_s=0$ and $m_s=\pm1$, respectively.  This definition avoids both the spin-projection noise (since the initial states are always assumed to be eigenstates) and the spin-dependent signal variance, since the total variance for the difference between two uncorrelated measurements is
\begin{equation}
\sigma_\mathrm{diff}^2 = \sigma_0^2+\sigma_1^2.
\end{equation}
This differential SNR measure is a useful figure-of-merit for many types of experiments with NV spins, including Rabi oscillations, spin-coherence measurements, and Ramsey or Echo-based magnetometry experiments.

\setlength{\extrarowheight}{8pt}
\begin{table}
\caption{Best-fit parameters corresponding to the multi-SCC measurements in Fig.~4(c) of the main text.}
\begin{tabular}{|l|l|l|}
\hline
NV$^0$ init & $k_{35}$ & $k_{45}$ \\ \hline
$0.04\pm0.013$ & $0.033\pm0.07$ & $0.25\pm0.04$ \\\hline\hline
P$_{\text{sing}}$ & $k_{51}/k_{52}$ & $m_s=0$ init \\\hline
$0.32\pm0.04$ & $2.26\pm0.01$ & $0.85\pm0.06$ \\
\hline
\end{tabular}
\label{tab:SCCFit}
\end{table}

\subsection{Ideal thresholding}
To begin, we analyze the SNR of an ideal single-shot thresholding condition, i.e., assuming perfect charge-state readout but imperfect SCC efficiency, in which we assign the outcome of NV$^-$ to 1, and NV$^0$ to 0. The differential signal is then defined as the difference in NV$^-$ outcomes for initial preparation in $m_s=0$ or $m_s=1$:
\begin{align}
\braket{S_\mathrm{threshold}} = \beta_0 - \beta_1,
\end{align}
where $\beta_i$ is the probability of detecting NV$^-$ given an initial spin state $m_s=i$. The variance associated with each spin state is given by
\begin{subequations}
\begin{align}
\sigma_{SCC}^2 &  = \braket{S^2}-\braket{S}^2 \\
\sigma_{SCC, i}^2 & = (0)^2(1-\beta_i) + (1)^2(\beta_i) - (\beta_i)^2 \\
 & = \beta_i(1-\beta_i).
\end{align}
\end{subequations}
Thus the total signal to noise ratio is
\begin{align}
\mathrm{SNR}_\mathrm{threshold} = \frac{\beta_0 - \beta_1}{\sqrt{\beta_0(1-\beta_0) + \beta_1(1-\beta_1)}}.
\end{align}
We use this expression to calculate the maximum SNR in the main text using the values of $\beta_i$ extracted from our SCC measurements. This formulation also allows us to calculate the single-shot readout fidelity of the electron spin by assigning the outcome of NV$^-$ to signify the spin state of $m_s=0$ and NV$^0$ to signify $m_s=\pm1$. The fidelity is calculated by finding the mean error of both readouts,
\begin{align}
\epsilon_0 = P(\text{NV}^0|m_s=0) = 1-\beta_0\\
\epsilon_{\pm1} = P(\text{NV}^-|m_s=\pm1) = \beta_1 \\
\epsilon_{total} = \frac{1}{2}(\epsilon_0 + \epsilon_{\pm1}). 
\end{align}
The readout fidelity, corresponding to the degree of confidence in determining the spin, is then given by 
\begin{align}
\mathcal{F}_s = 1 - \epsilon_{total}=\frac{1}{2}(1+\beta_0-\beta_1). \label{eq:Fidelity}
\end{align}

\subsection{Accounting for photon shot noise}

As we will see, in some cases it is beneficial to reduce the readout time even at the expense of lower-fidelity charge-state readout.  This is particularly true when the SCC efficiency is poor, and repeated averaging can help to reduce the uncertainty associated with the underlying random binomial process. In this case, we need to account for the shot-noise introduced in the charge-state readout step. This procedure is similar to that of traditional PL readout, except the underlying photon distribution function we are sampling is not Poissonian, but a joint distribution function accounting for the two Poissonian distributions corresponding to the NV$^-$ and NV$^0$ charge states, and the random SCC process.

Our measurement in this case is the number of photons detected rather than a binary outcome, but as before we define the signal as the difference in the mean number of photons $\alpha_i$ detected given an initial spin state $i$,
\begin{align}
\braket{S_{photon}} = \alpha_0 - \alpha_1.
\end{align}
There are two random processes occurring in this readout: the probabilistic SCC mechanism (random variable $y$) which is characterized by the efficiency parameters, $\beta_i$, and emission of a random number of photons based on the outcome of the charge state conversion (random variable $x$). This requires a joint probability density function given by
\begin{equation}
f_{X,Y}(x,y) = \left \{ \begin{aligned}
\frac{\eta_0^xe^{-\eta_0}}{x!}(1-\beta_i),\qquad  x=0,1,2...,\quad y=0 \\
\frac{\eta_-^xe^{-\eta_-}}{x!}\beta_i,\qquad  x=0,1,2...,\quad y=1
\end{aligned}
\right .
\end{equation}
where $\eta_0$ and $\eta_-$ correspond to the mean number of photons detected from the neutral and negative charge states, respectively. From this distribution, we can calculate the mean number of photons detected for a given charge state readout duration and SCC efficiency:
\begin{align}
E(X) = \sum_y E(X|Y=y)P(Y=y),
\end{align} 
which is interpreted as being the expected number of photons detected given either a successful or unsuccessful spin-to-charge conversion event, weighted by the probability of that event occurring. The resulting mean signals are
\begin{subequations}
\begin{align}
\alpha_0 = \beta_0\eta_- + (1-\beta_0)\eta_0, \\
\alpha_1 = \beta_1\eta_- + (1-\beta_1)\eta_0,
\end{align}
\end{subequations}
in agreement with basic intuition about the underlying weighted Poisson processes.

The variance is determined in a similar manner.  From the following calculation,
\begin{subequations}
\begin{align}
E(X^2) = \sum_y E(X^2|Y=y)P(Y=y) \\
= (1-\beta_i)\sum_xx^2\frac{\eta_0^xe^{-\eta_0}}{x!} + \beta_i\sum_xx^2\frac{\eta_-^xe^{-\eta_-}}{x!} \\
 = (1-\beta_i)(\eta_0^2 + \eta_0) + \beta_i(\eta_-^2 + \eta_-),
\end{align}
\end{subequations}
we calculate the noise associated with each spin state as 
\begin{align}
\sigma^2_i = E(X^2) - E(X)^2.
\end{align}
The total single-shot SNR is then given by
\begin{align}
\textrm{SNR}_\mathrm{S.S.} = \frac{\alpha_0 - \alpha_1}{\sqrt{\sigma^2_0 + \sigma^2_1}}.\label{eq:SNRSS}
\end{align}

The above formulation fully accounts for both the probabilistic SCC process and shot noise in the charge-state readout. We can further extend it to include time-averaged measurements by noting that in both cases the noise is reduced by $N^{-1/2}$ where $N$ is the number of repeats, implying that the total SNR will increase as $N^{1/2}$. This becomes advantageous particularly when the SCC efficiency is poor, when it can be better to average several measurements with short readout times (sometimes, even measurements where $<$1 photons are detected in each shot) than to perform a single measurement with high charge-state fidelity.

\subsection{Spin readout noise}

Several previous works focusing on magnetometry \cite{Shields2015,Lovchinsky2016} use the spin-readout noise, $\sigma_R$, as a figure of merit to characterize the performance of their measurements.  This is meant to account for both the spin-projection noise resulting from a measurement of spin-superposition states as well as experimental imperfections, where $\sigma_R=1$ corresponds to the standard quantum limit.

To connect our results to those experiments, we return to the thresholding case and calculate the average expected signal including spin populations, given by
\begin{align}
\braket{S} = \cos^2\left(\frac{\theta}{2}\right)\beta_0 + \sin^2\left(\frac{\theta}{2}\right)\beta_1,
\end{align}
where the angle $\theta$ determines the spin populations through $p_0=\cos^2(\theta/2)$. The spin readout noise is then defined as the point at which the noise equals the change in the signal as a function of phase, given by:
\begin{widetext}
\begin{align}
\sigma_R = \frac{\sigma_S}{\frac{\partial \braket{S}}{\partial\theta}}  
= \frac{\sqrt{(2-\beta_0-\beta_1-(\beta_0-\beta_1)\cos\theta)(\beta_0+\beta_1+(\beta_0-\beta_1)\cos\theta)}}{|\beta_1 - \beta_0|\sin\theta}.
\end{align}
\end{widetext}
Whereas the spin readout noise is typically defined for equal initial spin populations ($\theta=\pi/2$), note that this expression depends on the underlying spin population as long as $\beta_0\neq\beta_1$, and the minimum does not necessarily occur at $\theta=\pi/2$.  This implies that it might be beneficial to choose magentometry pulse sequences that set a different operational point for maximum performance.

\subsection{Practical Performance Considerations}
\begin{figure}
\includegraphics[width=\linewidth]{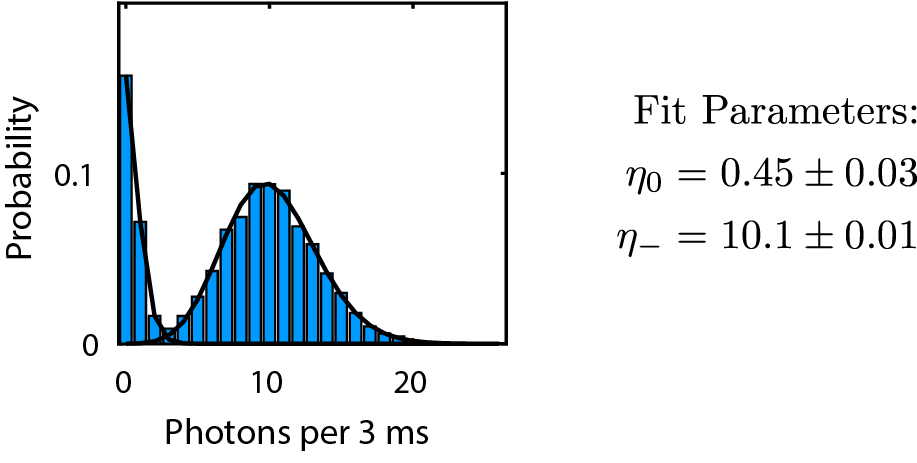}
\caption{Experimental results for the optimized charge state readout with along with the fit parameters determining the mean counts for each charge state.}
\label{fig:CSParams}
\end{figure}
We optimized our charge state readout to gain the highest signal to noise.  The resulting photon distribution following green initialization is shown in Fig.~\ref{fig:CSParams}. For a \SI{3}{\milli\second} readout at \SI{220}{\nano\watt}, we determine the average number of photons per measurement to be $\eta_0=0.45$ and $\eta_-=10$. These values are determined from the fit of the photon histogram to the sum of two Poisson distributions, where the two means correspond to the average number of counts. We use the optimized SCC efficiency from the main text ($\beta_0=0.80\pm0.02$ and $\beta_- = 0.60\pm0.02$) corresponding to 10 repeats. These parameters are used in the expressions derived above to quantify the spin-readout performance with regards to SNR, single-shot fidelity, and spin readout noise as quoted in the main text. The ideal case assuming 100$\%$ ionization (but still imperfect spin initialization) was determined from the best-fit parameters in the master-equation model, and corresponds to SCC efficiency of $\beta_0^{ideal}=0.7$ and $\beta_1^{ideal}=0.19$.  Assuming perfect spin initialization, these values should approach the underlying $\sim$10:1 spin-dependent branching ratio of the excited-state ISC, yielding $\beta_0=0.9$, $\beta_1=0.1$ and SNR=1.9.  We believe there might be ways to improve the spin purity in the future, possibly even using SCC, but since this is somewhat speculative we have not included these idealized estimates in Fig.~4 of the main text.

\section{Time-averaged measurements}

Whereas longer (highest-fidelity) charge-state readout is best for single-shot measurements, the overall performance of time-averaged measurements depends non-trivially on the duration of the charge state readout, and implicitly on the intensity of the readout laser.  By moving to shorter readout times, more iterations can be averaged, which reduces the noise associated with the SCC process.  Furthermore, at shorter readout times we can use a higher-intensity readout laser to increase the underlying count rates, as long as we maintain the strong contrast between the NV$^-$ and NV$^0$ charge states.

To account for these variations, the expected photon collection rates for the the bright and dark state can be written in the form:
\begin{subequations}\label{eq:RateParameterization}
\begin{align}
\gamma_- = c\Gamma_{\textrm{sat}}\frac{1}{1 + \frac{P}{P_{\textrm{sat}}}} + \gamma_{\textrm{Det.}} \\
\gamma_0 = D\frac{P}{P_{\textrm{sat}}} + \gamma_{\textrm{Det.}}
\end{align}
\end{subequations}
Where $c$ is the photon collection efficiency (0.005 for our setup), $\Gamma_{\textrm{sat}}$ is the theoretical maximum collection rate from NV$^-$ ($\sim$\SI{50}{\mega\hertz}), $P/P_{\textrm{sat}}$ is the scaled power, $D$ is the background scaling of the dark state, and $\gamma_{\textrm{Det.}}$ is the detector dark-count background (\SI{20}{\hertz}).  

To determine the optimium power setting, we note that the ionization rate, $\Gamma_\mathrm{ion}$, scales quadratically with power \cite{Aslam2013}, and so as the readout time, $\tau_R$ changes, we can scale the power with $P\sim 1/\sqrt{\tau_R}$ in order to maintain the condition that $\Gamma_\mathrm{ion}<1/\tau_R$, as required for strong PL contrast between the the charge states.  We can therefore make the following replacement in the expressions (\ref{eq:RateParameterization}):
\begin{equation}
\frac{P}{P_\mathrm{sat}}=\sqrt{\frac{\tau_{R0}}{\tau_R}}, \label{eq:PowerVariation}
\end{equation}
where $\tau_{R0}$ is the readout time at which we would reach saturation power.  From our optimized high-fidelity readout from Fig.~\ref{fig:CSParams}, at a power well below saturation producing $\gamma_-=3.37$ \si{\kilo\counts\per\second}, we calculate $\tau_{R0}=550$ \si{\nano\second}. We also characterize the NV$^0$ background scaling from the same measurement, since we know the count rate at that same power is \SI{.15}{\kilo\counts\per\second}. 

\subsection{Speed Up Calculations}
Using the above expressions, we can calculate the optimized single-shot SNR for the SCC-readout technique for any readout time, accounting for photon shot noise. The anticipated mean signals are simply
\begin{subequations}
\begin{align}
\eta_0 = \gamma_0\tau_R \\
\eta_- = \gamma_-\tau_R
\end{align}
\end{subequations}
Note that the assumption of two independent Poisson distributions underlying the photon arrival statistics is still valid since we are ensuring through eqn.~(\ref{eq:PowerVariation}) that no charge switching events occur during the readout. 

\begin{figure}
\includegraphics[width=\linewidth]{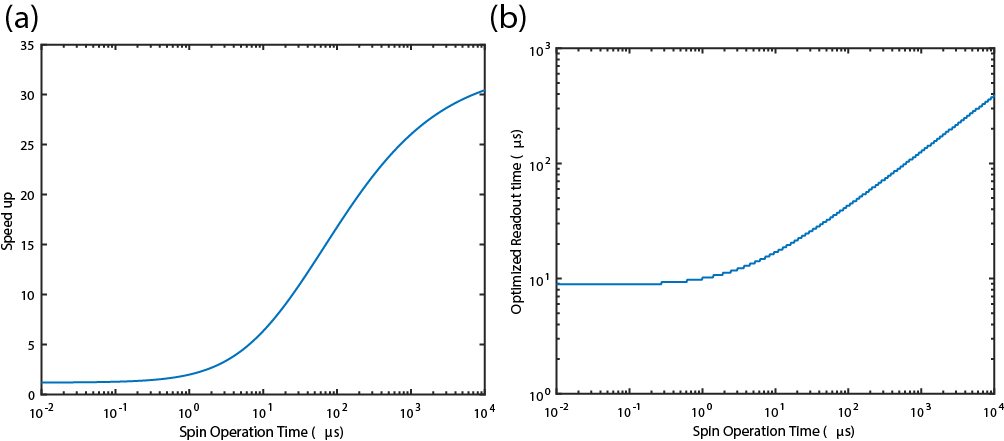}
\caption{Results of Speed Up Optimization (a) Resulting speed up plotted versus the spin operation time. (b) Required charge readout time to achieve the speed up in (a). Note the shortest readout time, \SI{95}{\micro\second}, is much shorter than the anticipated value for operating at the saturation value, thus we are still working far from saturation fluorescence values, yet maintaining the contrast.}
\label{fig:speedup}
\end{figure}

To compare the time-averaged signals resulting from SCC readout with traditional PL, we numerically optimize the readout time to find the minimum total integration time, $T$, required to achieve a time-averaged SNR=1, which is given by:
\begin{align}
T = \frac{\tau_I + \tau_O + \tau_R}{\textrm{SNR}_\mathrm{S.S.}^2},
\end{align}
where $\tau_I=1$~\si{\micro\second} is the initialization time, $\tau_O$ is the operation time on the spin (time between the initialization and measurement), $\tau_R$ is the charge readout time, and SNR$_\mathrm{S.S.}$ is given by eqn.~(\ref{eq:SNRSS}) for SCC readout and by the corresponding expression,
\begin{equation}
\mathrm{SNR}_\mathrm{S.S., PL}= \frac{\alpha_0-\alpha_1}{\sqrt{\alpha_0+\alpha_1}},
\end{equation}
for the case of traditional PL. Note that for traditional PL readout, the single-shot SNR is constant as a function of operation time, since the readout time and corresponding count rates are fixed (we assume $\tau_{R,\mathrm{PL}}=200$~\si{\nano\second} and $\alpha_1=0.7\alpha_0$), however for SCC the SNR changes dramatically as a function of readout time, and we need to find the optimum for a given $\tau_O$. These values are plotted in Fig.~4(e) of the main text for both the demonstrated and ideal (full ionization) SCC processes along with PL readout for our device.  Fig.~\ref{fig:speedup} shows the optimized speedup, given by
\begin{align}
\textrm{Speedup} = \frac{T_\mathrm{PL}}{T_\mathrm{SCC}},
\end{align}
for our demonstrated SCC process, along with the optimized charge-state readout time.  Note that for $\tau_O<1$~\si{\micro\second}, $\tau_R$ plateaus around \SI{10}{\micro\second}, which is still far larger than the point $\tau_{R0}=550$~\si{\nano\second} at which saturation effects would play a role.

\section{Alternative readout comparisons}
In the main text, we report the SNR of other enhanced readout techniques from the literature, measured relative to their reported PL readout. In order to do this, we converted all measurements to differential SNR as outlined below. Refs~\cite{Shields2015,Lovchinsky2016} use a similar metric (which is an extension of the method originally developed by Jiang \textit{et al.} \cite{Jiang2009}), namely the spin readout noise and the ``fidelity'', $\mathcal{F} = \sigma_R^{-1}$ (note, this is not equivalent to the single-shot readout fidelity parameter from eqn.~(\ref{eq:Fidelity})). We can relate these quantities directly to the SNR by noting that the definition of spin readout noise for PL techniques quoted in those works is related to the SNR as
\begin{subequations}
\begin{align}
\sigma_R = \sqrt{1 + \frac{2(\alpha_0 + \alpha_1)}{(\alpha_0 - \alpha_1)^2}} \\
\sigma_R = \sqrt{1 + \frac{2}{\text{SNR}^2}}
\\
\rightarrow \text{SNR} = \frac{\sqrt{2}}{\sqrt{\sigma_R^2 - 1}}
\end{align}
\end{subequations}

\begin{table*}
\caption{Comparison of enhanced spin-readout techniques.}
\begin{tabular*}{\textwidth}{p{3cm} p{1.4cm} p{1.5cm} p{2cm} p{2cm} p{6.5cm}}
\textbf{Study} & \textbf{S.S. SNR} & \textbf{SNR gain} & \textbf{Optimal SNR} & \textbf{Saturation count rate} & \textbf{Requirements} \\
\hline
Lovchinsky \textit{et al.} \cite{Lovchinsky2016} & 0.29 &  7.2 & ? &  \SI{170}{\kilo\counts\per\second} & Strongly coupled nuclear spin, precise magnetic field alignment, 3 microwave sources \\

Shields \textit{et al.} \cite{Shields2015} & 0.55 & 4.2 & $\sim$0.55 & \SI{950}{\kilo\counts\per\second} & 3 illumination colors, single-shot charge readout$^\ast$ \\

Robledo \textit{et al.} \cite{Robledo2011} & 3.5$^\dag$ & N/A & ? & \SI{1}{\mega\counts\per\second} & Cryogenic ($<$\SI{10}{\kelvin}) temperatures, high collection efficiency \\

This work & 0.32 & 5.8 & 0.84 & \SI{250}{\kilo\counts\per\second} & 3 illumination colors, charge state \textit{contrast}, applicable to ensembles \\
\hline
\multicolumn{6}{l}{$^\ast$\footnotesize{As demonstrated, although this SCC approach could also benefit from multishot averaging.} }\\
\noalign{\vskip -6pt}  
\multicolumn{6}{l}{$^\dag$\footnotesize{Value includes enhanced spin-initialization purity.}}\\
\end{tabular*}
\label{table:spinreadout}
\end{table*}

Lovchinsky \textit{et al.} \cite{Lovchinsky2016} report a standard fluorescence readout fidelity of $\mathcal{F}=0.03$, which corresponds to an SNR=0.04. Their optimized readout has a fidelity $\mathcal{F}=0.2$ which corresponds to an SNR=0.29. This corresponds to a 7.25 increase in the spin readout SNR. While they do not state their exact collection efficiency, we can estimate it using their reported PL fidelity of 0.03, solving for $\alpha_0$, and assume a 30$\%$ PL contrast (i.e. $\alpha_1=0.7\alpha_0$). A further assumption of a \SI{200}{\nano\second} readout duration puts their peak count rate of \SI{170}{\kilo\counts\per\second}.

Shields \textit{et al.} report a standard fluoresence readout noise level of $\sigma_R^{PL}=10.6$, corresponding to an SNR=0.13. Their enhanced technique directly quotes the spin-to-charge conversion efficiency, of $\beta_0=0.16$ and $\beta_1=0.5$, corresponding to an SNR=0.55. This corresponds to a 4.2 increase in the spin readout SNR. Based on their reported $\beta_i$ SCC parameters, Shields \textit{et al.} have optimized their technique for maximum SCC efficiency via ionization of the triplet state, and, in contrast to singlet-SCC, there is no possibility for further gains through repeated application of the procedure since the spin state is already destroyed after one cycle of the protocol.  The peak count rate is reported in the first figure, and is an impressive \SI{950}{\kilo\counts\per\second}.  

While not a direct comparison to these room-temperature techniques, we also consider the performance of the single-shot readout technique demonstrated by Robledo \textit{et al.} \cite{Robledo2011} using coherent optical transitions at cryogenic temperatures. This technique takes advantage of the excited state fine structure to directly probe spin selective transitions. It provided an SNR=3.5, but more importantly a single-shot fidelity of 93$\%$. They report a peak count rate of $\sim$\SI{1}{\mega\counts\per\second}, according to their supporting information.

In the present work, our SIL-enhanced device yields a single-shot SNR=0.055 for traditional PL measurements, based on detecting 0.05 and 0.034 photons for $\alpha_0$ and $\alpha_1$ respectively. Our singlet-based multi-SCC technique has a maximum SNR= 0.32, using the thresholding expression. This corresponds to a 5.8-fold increase in the spin-readout SNR. The optimized scenario of 100\% singlet ionization should produce an SNR=0.85, corresponding to a 15-fold boost over PL.  The upper limit of the technique, also assuming perfect spin initialization, is set by the $\sim$10:1 branching ratio of the excited-state ISC, which yields SNR=1.9 and a single-shot fidelity, $\mathcal{F}=0.9$.

The results of these comparison, along with the requirements for each technique, are summarized in Table~\ref{table:spinreadout}.

\bibliographystyle{apsrev4-1}
\bibliography{ionization}